\documentclass[runningheads]{llncs}

\usepackage[year=2026,ID=10887]{eccv}

\usepackage{eccvabbrv}

\usepackage{graphicx}
\usepackage{booktabs}

\newif\ifdraft
\drafttrue

\ifdraft

\definecolor{silviablue}{HTML}{1D4F91}

\definecolor{LavenderBlue}{rgb}{0.69,    0.87,    0.54}
\usepackage{tabularx}
\usepackage{booktabs}
\usepackage{makecell}
\usepackage{longtable}
\usepackage[table]{xcolor}
\definecolor{white}{rgb}{1,1,1}
\definecolor{lightbluishgrey}{rgb}{0.76471,0.84824,0.91647}
\newcommand{\basemesh}{\mathcal{M}_b}
\newcommand{\restmesh}{\mathcal{M}_r}
\newcommand{\baseverts}{\mathcal{V}_b}
\newcommand{\basefaces}{\mathcal{F}_b}
\newcommand{\highlightedfaces}{F^*}
\newcommand{\highlightedmesh}{\mathcal{M}_b^*}
\newcommand{\highlightednormal}{\mathbf{n}^*}
\newcommand{\objectmesh}{\mathcal{M}_o}
\newcommand{\objectverts}{\mathcal{V}_o}
\newcommand{\objectfaces}{\mathcal{F}_o}
\newcommand{\transformation}{T}
\newcommand{\rigid}{R}
\newcommand{\parameters}{\theta}
\newcommand{\rotation}{\mathbf{e}_1, \mathbf{e}_2}
\newcommand{\translation}{\mathbf{t}}
\newcommand{\scale}{s}
\newcommand{\vertex}{v}
\newcommand{\vertext}{t}

\newcommand{\thresholdprox}{\ell}
\newcommand{\gaussian}{\mathcal{N}}
\newcommand{\uniform}{\mathcal{U}}
\newcommand{\numnonintersecting}{n}
\newcommand{\numtimesteps}{m}
\newcommand{\contactthreshold}{h}
\newcommand{\tr}{tr}

\newcommand{\ourmethod}{MeshOn}
\usepackage{mathtools}
\usepackage[table]{xcolor}
\usepackage{colortbl}
\definecolor{headercolor}{RGB}{230,230,230}
\definecolor{zebra}{RGB}{245,245,245}

\usepackage[accsupp]{axessibility}

\usepackage{hyperref}
\usepackage{wrapfig}
\usepackage{orcidlink}

\begin{document}

\title{MeshOn: Intersection-Free\\Mesh-to-Mesh Composition}

\author{Hyunwoo Kim\inst{1,2} \and
Itai Lang\inst{2} \and Hadar Averbuch-Elor\inst{3} \and Silvia Sellán\inst{1} \and Rana Hanocka\inst{2}}
\authorrunning{H. Kim et al.}
\institute{Columbia University \and University of Chicago \and Cornell University}

\maketitle

\begin{center}
    \centering
    \includegraphics[width=\linewidth]{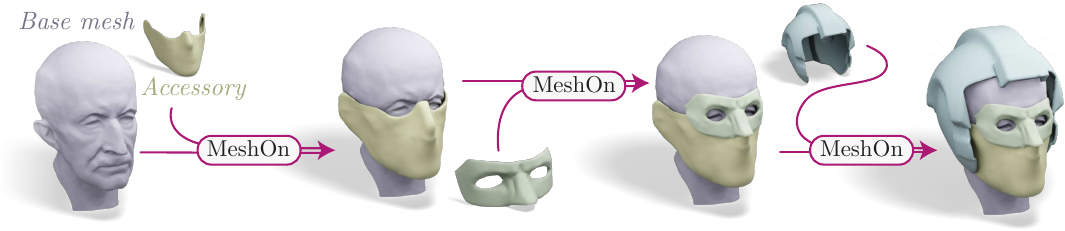}
    \captionof{figure}{\ourmethod{} is a multi-step optimization algorithm that fits accessories onto meshes realistically, tightly, and without intersections.}
    \label{fig:teaser}
\end{center}%

\begin{abstract}
We propose \ourmethod{}, a method that finds physically and semantically realistic compositions of two input meshes. Given an accessory, a base mesh with a user-defined target region, and optional text strings for both meshes, \ourmethod{} uses a multi-step optimization framework to realistically fit the meshes onto each other while preventing intersections.
We initialize the shapes' rigid configuration via a structured alignment scheme using Vision-to-Language Models, which we then optimize using a combination of attractive geometric losses, and a physics-inspired barrier loss that prevents surface intersections. 
We then obtain a final deformation of the object, assisted by a diffusion prior.
Our method successfully fits accessories of various materials over a breadth of target regions, and is designed to fit directly into existing digital artist workflows. We demonstrate the robustness and accuracy of our pipeline by comparing it with generative approaches and traditional registration algorithms. Our project page is at \hyperlink{https://threedle.github.io/MeshOn/}{https://threedle.github.io/MeshOn/}.
\end{abstract}

\section{Introduction} \label{sec:introduction}

\emph{Mesh composition} is the process of assembling several pre-modeled 3D objects into a unified asset.
It is a fundamental part of many 3D content creation workflows, in which an artist selects \emph{accessories} from an existing library and manually rotates, translates and deforms them to fit realistically on a character's \emph{base mesh} without intersections (see \autoref{fig:teaser}).
Automating this tedious task is a critical and challenging research question combining \emph{geometric} and \emph{physical} constraints (the two meshes must be close to each other without intersecting) with \emph{semantic} alignment (e.g., a pair of glasses should sit on the eyes in the correct orientation).

\begin{figure*}
    \centering
    \vspace{3mm}
    \includegraphics[width=\textwidth]{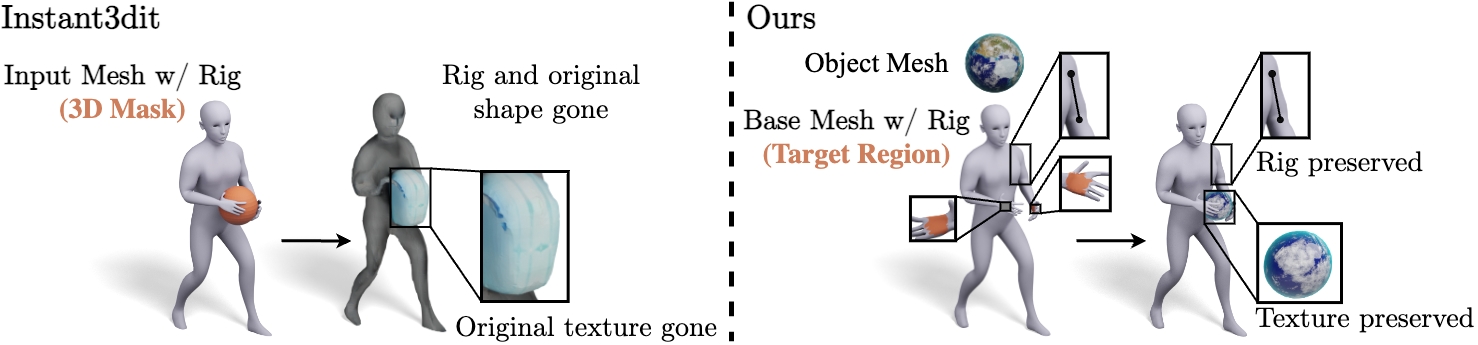}
    \captionof{figure}{Digital artists often load shapes from pre-existing libraries of meshes that include additional information like animation rigs, skinning weights, texture maps and more. Generative shape editing methods like Instant3dit \cite{barda2025instant3dit} discard this information, performing global changes to the shapes and merging them together. \ourmethod{} is designed to fit exactly within this common artistic pipeline, and perfectly preserves all pre-existing information on the input meshes.}
    \label{fig:instant3dit_comparison}
    \vspace{-5.9mm}
\end{figure*}%

However, recent advances in 3D content creation have focused mostly on \emph{generative} tasks; for example, creating shapes from text inputs~\cite{lin2023magic3d}.
These methods produce doubtlessly impressive results; yet, they do so at the cost of artist control and by relying on implicit geometric representations (e.g., radiance fields~\cite{mildenhall2021nerf} or Gaussian splats~\cite{kerbl20233d}) that are not directly usable in graphics pipelines.
Even when triangle meshes are generated~\cite{hao2024meshtron}, these are significantly lower in quality than those already available to artists in pre-existing shape libraries, lacking critical attributes like textures, animation rigs, and distinct material parameters.

We propose \ourmethod{}, an optimization framework that progressively rotates, translates and then deforms an accessory until it fits tightly onto a base mesh while maximizing geometric, physical and semantic alignment.
We model this task's complexity by combining geometric losses based on distance between the meshes with physical losses preventing intersections and diffusion-based semantic guidance, and
devise a unique, multi-step optimization strategy in which each individual loss is introduced sequentially.

To succesfully navigate the ambiguities of the composition problem, we first assume that some coarse segmentation of the local ``fit'' region is highlighted, and use Vision-to-Language Model~\cite{gemini} agents to ensure adequate initialization. 
From this semantically-guided initial configuration, our algorithm drives the accessory closer to the base mesh, while using a GPU-optimized Bounding Volume Hierarchy to efficiently enforce non-intersection.
Similarly to how an artist may refine a pre-modeled accessory after optimizing its coarse position, our method optimizes a final deformation of the object through its Jacobians~\cite{aigerman2022neural}, maintaining physical realism by accounting for its (given or inferred) material parameters.

In this paper, we present a geometric optimization framework for reliably composing 3D meshes with accessories.
Through qualitative and quantitative comparisons, we show that our geometric optimization outperforms more general generative and 3D registration approaches for the task of mesh composition, while remaining intersection-free.
We show the robustness and applicability of our method on a diverse set of base meshes and accessories inspired by 3D content generation pipelines.

\begin{figure*}[t]
    \centering
    \vspace{3mm}
    \includegraphics[width=\textwidth]{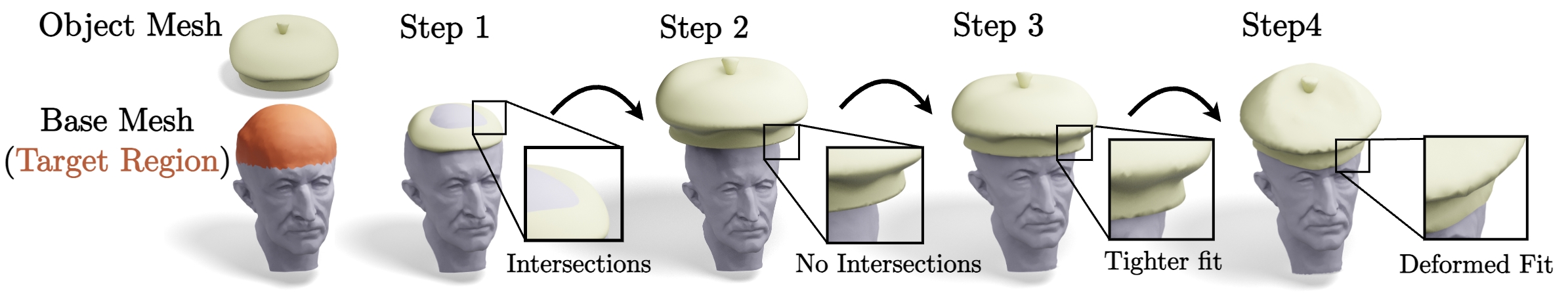}   
    \caption{We compose the two meshes through a multi-step optimization pipeline: from an initialization obtained with a Vision-to-Language Model (\autoref{sec:initialization}), we start by obtaining a tight fit containing intersections (\autoref{sec:step1}), which we then resolve (\autoref{sec:step2}). After finetuning the fit to obtain the best possible rigid fit (\autoref{sec:step3}), we improve it further by allowing small deformations in the accessory object (\autoref{sec:step4}).}
    \label{fig:main_method}
    \vspace{-3mm}
\end{figure*}

\section{Related Work} 

\label{sec:related_work}

\subsection{3D Registration}

The alignment of 3D objects is a fundamental problem in computer vision and graphics that has been studied thoroughly for decades \cite{mckay2003review,salvi2007survey, tam2013registration}. In this task, one seeks to find a transformation of an object to minimize an error metric defined with respect to another. A prominent work in this field is the Iterative Closest Point \cite{besl1992method,segal2009generalized}, which finds the optimal rotation and translation between two point clouds to minimize the distance between them. Among follow-up works to ICP are the globally optimal ICP (Go-ICP) \cite{yang2015go} and Fast Global Registration (FGR) \cite{zhou2016fast}.

In the deep learning era, numerous works have leveraged the power of neural networks for the registration problem, where the Euclidean distance between points has been replaced with the similarity of points in a learned feature space \cite{litany2017deep,aoki2019pointnetlk,wang2019dcp,huang2020fmr,choy2020dgr,lang2021dpc,yew2022regtr,jiang2023se}. Our mesh composition problem can be thought of as a registration problem between the asset mesh and the corresponding region on the body mesh. However, different from previous work that aligned two observations of the \textit{same} object, the asset is completely different from the target body region, making the problem highly challenging.

\subsection{Mesh Deformation}
A large body of research has been dedicated to surface deformation. Classical methods typically define an energy function subject to which the mesh is optimized, where the objective embodies desired properties of the deformed result \cite{yu2004poisson,botschsorkine08deformsurvey,skinningcourse:2014,Fulton:LSD:2018}. As-Rigid-As-Possible \cite{sorkine2007rigid} and Laplacian surface editing \cite{sorkine2004laplacian} are seminal works in this domain, which encourage smooth and rigid deformations. Such methods are applied to the subject mesh independently, without considering its interaction with other objects.

In light of the unprecedented success of deep learning, researchers have incorporated deep foundation models as implicit guidance for the deformation process instead of the traditional hand-crafted optimization objective, leveraging the rich semantic priors embodied in these models and enjoying their intuitive language-based interface \cite{michel2022text2mesh,xu2024fusiondeformer,yang2024dreammesh,chen2024textmeshrefine}. TextDeformer \cite{michel2022text2mesh} used the CLIP model for deforming a source mesh toward a textual description, like turning a hand into an octopus. Follow-up works employed a powerful diffusion model for other deformation-driven creative applications, including concept blending \cite{kim2025meshup} and identity-preserving geometric stylization~ \cite{dinh2025geometry}. 

Similarly, we also leverage a diffusion model to guide the deformation of the accessory mesh. In contrast to prior work, the deformation in our case is constrained by the body mesh it should be composed onto, considering both their global semantic alignment and their local geometric fitting.

A recent work tackled the problem of designing creative 3D objects that fit a body mesh \cite{guo2025craft}. However, in stark contrast to our work, in CRAFT \cite{guo2025craft}, the contact points between the meshes are \textit{given} as input, while our method \textit{finds} the alignment between the meshes from a \textit{random initialization} of the asset object. 

More relevant to our work is Instan3dit \cite{barda2025instant3dit}. In this work, the authors augment a base mesh with an accessory, such as adding a honey pot between the hands of a bear. The edit is implemented as a multi-view inpainting problem, followed by a 3D reconstruction from the posed images. While the results are visually appealing, the contact between the local edit and the body mesh is not modeled explicitly, leading to the gluing of the two. In our work, on the other hand, we consider the mesh interaction explicitly and successfully preserve the base and asset mesh properties (see \cref{fig:instant3dit_comparison}).

\section{Method} 
\label{sec:method}
\vspace{-2mm}
The inputs to our method are a base mesh $\basemesh = (\baseverts, \basefaces)$ (e.g., a bust, as shown in \autoref{fig:main_method}) and an object or accessory mesh $\objectmesh = (\objectverts, \objectfaces)$ (e.g., a hat) in arbitrary relative position, together with a highlighted region of the base mesh $\highlightedmesh \subseteq \basemesh$ (the orange region in \autoref{fig:main_method}) and an optional text string describing both shapes.
The algorithm we will detail in this section will output a transformed object mesh $\transformation (\objectmesh) = (\transformation (\objectverts), \objectfaces)$ that can be superimposed with the base mesh \emph{tightly}, in a \emph{semantically and physically realistic} way \emph{without intersections}.

\begin{figure*}[t]
    \centering
    \vspace{3mm}
    \includegraphics[width=\textwidth]{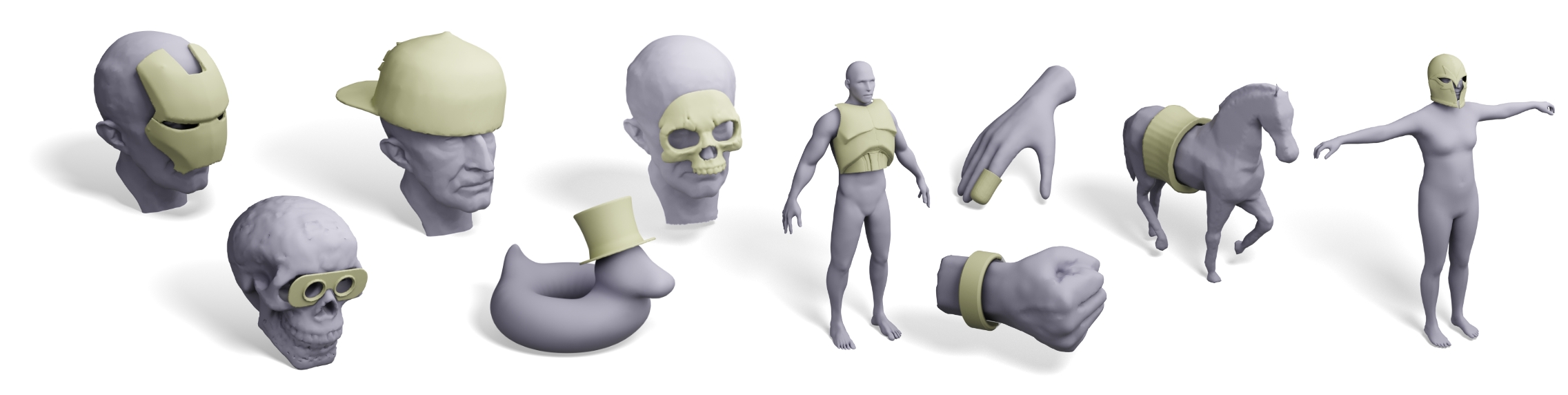}
    \captionof{figure}{\ourmethod{} is capable of fitting a wide variety of accessories on a range of different meshes. Our method handles challenging spatial relationships such as sliding glasses along a head to align with the ears, positioning hats and helmets to conform to head curvature, wrapping bands around articulated limbs, and fitting objects along long, curved surfaces. These examples demonstrate the flexibility of our fitting trajectory and its ability to adapt accessories to widely varying shapes and poses.}
    \label{fig:gallery}
    \vspace{-3mm}
\end{figure*}%

These goals often act in opposition to one another; for example, achieving a tighter fit will often produce intersections, while removing intersections will tend to produce a looser fit.
We address this challenge through a multi-step optimization framework in which the different constraints, objectives and degrees of freedom of the problem are introduced one at a time.
From an initial (scaled) rigid transformation $\rigid_0$ (we defer discussion of initialization strategies to \autoref{sec:initialization}), we begin by rigidly and tightly fitting the object onto the base mesh, potentially \emph{with} intersections (see $\rigid_1$ in \autoref{sec:step1}).
Then, we simulate possible rigid object attachment trajectories using physically-inspired losses to obtain a loose \emph{non-intersecting} configuration $\rigid_2$ (see \autoref{sec:step2}), which we then refine, first into a tight rigid non-intersecting fit $\rigid_3$ (see \autoref{sec:step3}) and then through a diffusion-guided deformation (see \autoref{sec:step4}) to produce our final transformation $\transformation$.

\subsection{Step 1: Tight fit with intersections}
\label{sec:step1}

We will represent the space of all possible scaled rigid transformations through rotation, translation and scale parameters $\rigid(\parameters) = \rigid (\rotation, \translation, \scale)$, where $\rotation$ are the standard continuous 6D representation of 3D rotations \cite{zhou2019continuity}.
To find a rigid transformation that aligns the object mesh with the highlighted region of the base mesh (potentially while intersecting with it), one could naively optimize the two-sided vertex-to-mesh total distance between the base and (highlighted) object meshes
\begin{equation*}
    \sum_{\vertex_{i} \in \objectmesh} d (\rigid(\parameters)\vertex_i, \highlightedmesh)^2 +  \sum_{\vertex_{j} \in \highlightedmesh} d(\vertex_j, \rigid (\parameters)\objectmesh)^2\, .
\end{equation*}
Optimizing such energy directly would present two key challenges. First, the quadratic computational complexity would place a severe limit on mesh size. Secondly, the dense computational graph would present storage and performance issues during GPU autodifferentiation.

We fix the first of these problems through a Bounding Volume Hierarchy tailor-made to exploit vectorized operations in the GPU.
We build our BVH in a similar way to a traditional Axis-Aligned Bounding Box hierarchy (see \cite{shirley2009fundamentals}, Chap. 12). However, we abort any top-down traversal when a box contains fewer than $K$ mesh triangles, resorting instead to a tensorized batch query in the GPU (in our experiments, we fix $K=1,000$).

To avoid overhead from dense computational graphs, and to make our algorithm robust to haphazardly defined contact regions, we introduce a \emph{masked} distance function $\hat{d}$ that is identical to the true distance $d$ but which zeroes out any entry above a threshold $\thresholdprox$ (in our experiments, $\thresholdprox=0.5$). This lets us define our \emph{proximity loss}
\begin{equation*}
    \mathcal{L}_{p}(\parameters) = \sum_{\vertex_{i} \in \objectmesh} \hat{d} (\rigid(\parameters)\vertex_i, \highlightedmesh)^2 +  \sum_{\vertex_{j} \in \highlightedmesh} \hat{d}(\vertex_j, \rigid (\parameters)\objectmesh)^2\, ,
\end{equation*}
which we optimize to find $\parameters^\star$ and, with it, our first scaled rigid transformation $\rigid_1 = \rigid (\parameters^\star)$ (see \autoref{fig:main_method}).

\subsection{Step 2: Trajectory-based intersection solve}
\label{sec:step2}

\begin{wrapfigure}[9]{r}{0.55\linewidth}
    \centering
    \vspace{-6mm}
    \includegraphics[width=\linewidth]{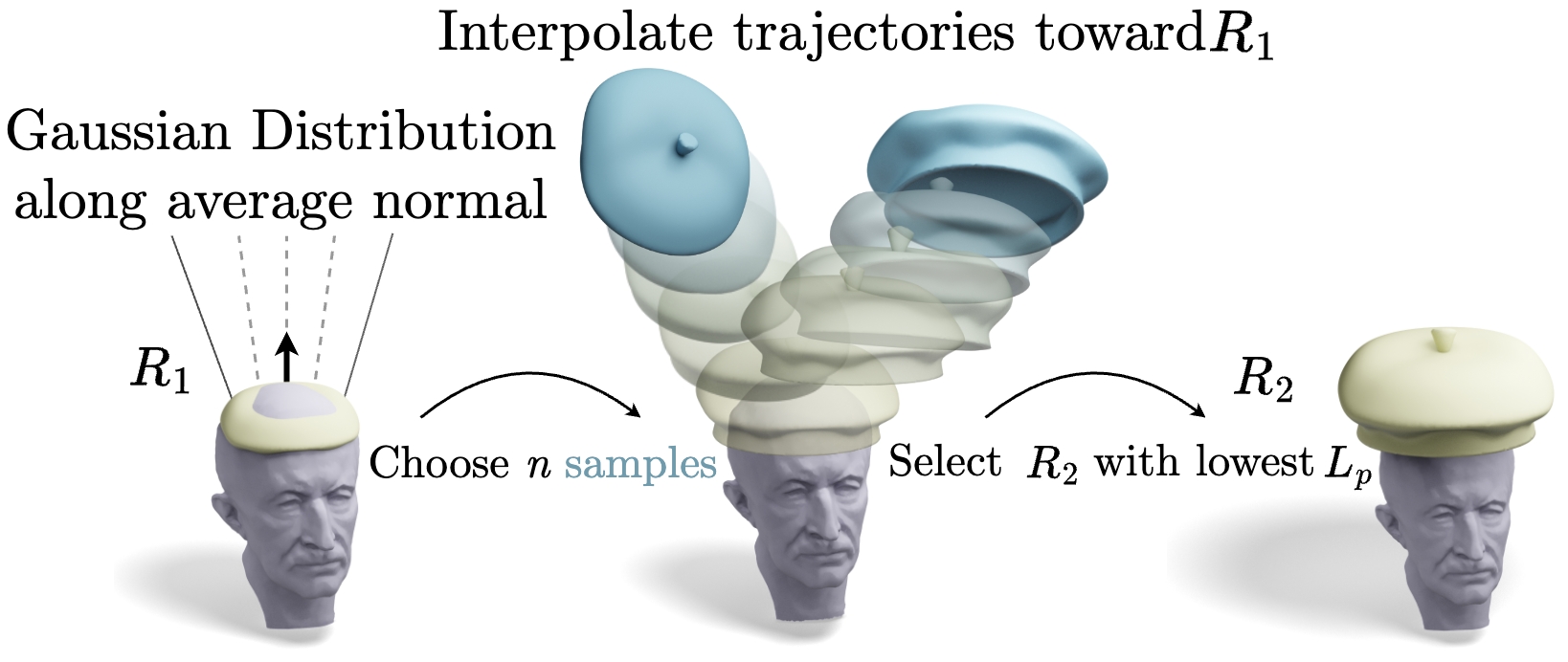}
    \caption{}
    \label{fig:trajectory_method}
\end{wrapfigure}

In the previous step, we rigidly transformed the object to tightly fit into the base mesh; however, this often introduces intersections between them (see \autoref{fig:trajectory_method}). We will now turn this intersecting configuration into a non-intersecting one, without compromising too much on tightness.

To do so, we will draw inspiration from how one puts on an accessory: for most tightly-fitting ones, there exists a valid, intersection-free trajectory that moves the object from a detached configuration into a tight fit with the base mesh (e.g., the process of putting on a hat or a ring).

We will begin by generating a large number of potential starting points for this trajectory, as shown in \autoref{fig:trajectory_method}. To do so, we draw rigid transformation's parameters from independent Gaussian and uniform distributions
\begin{align*}
\mathbf{e}_i \sim \gaussian (0,\sigma_{rot})\,,&\quad \scale \sim \uniform(\scale_{\min}, \scale_{\max})\,,\\
\translation_{\perp} \sim \gaussian (0,\sigma_{\perp})\,,&\quad \translation_{\parallel} \sim \gaussian (\mu_{\parallel},\sigma_{\parallel})
\end{align*}
where we have separated $\translation$ into the components orthogonal ($\translation_{\perp}$) and parallel ($\translation_{\parallel}$) to the average normal in the base mesh's highlighted region $\highlightednormal$ (see \autoref{fig:trajectory_method}, left). We draw samples from these distributions to produce candidate scaled rigid transformations $\rigid$ and check whether $\rigid \objectmesh$ intersects with the base mesh $\basemesh$, repeating this sampling process and progressively increasing the values of $\sigma_{rot}$ and $\sigma_{\parallel}$ until we find $\numnonintersecting$ non-intersecting candidate transforms $\rigid^1,\dots,\rigid^{\numnonintersecting}$ (in our experiments, $\numnonintersecting = 100$).

Once $\rigid^1,\dots,\rigid^{\numnonintersecting}$ have been identified, we use spherical linear interpolation (SLERP) to build candidate trajectories from each non-intersecting $\rigid^i$ to the intersecting transformation from the previous step $\rigid_1$ (see \autoref{fig:trajectory_method}, transparent frames). We sample $\numtimesteps$ discrete timesteps of each trajectory, keeping only the non-intersecting ones (in our experiments, $m=25$). After gathering all non-intersecting timesteps for all candidate trajectories $R^i$, we make $\rigid_2$ be the transformation with the lowest proximity loss $\mathcal{L}_{p}$ (\autoref{fig:trajectory_method}, right).

\begin{figure*}[t]
    \centering
    \vspace{3mm}
    \includegraphics[width=\textwidth]{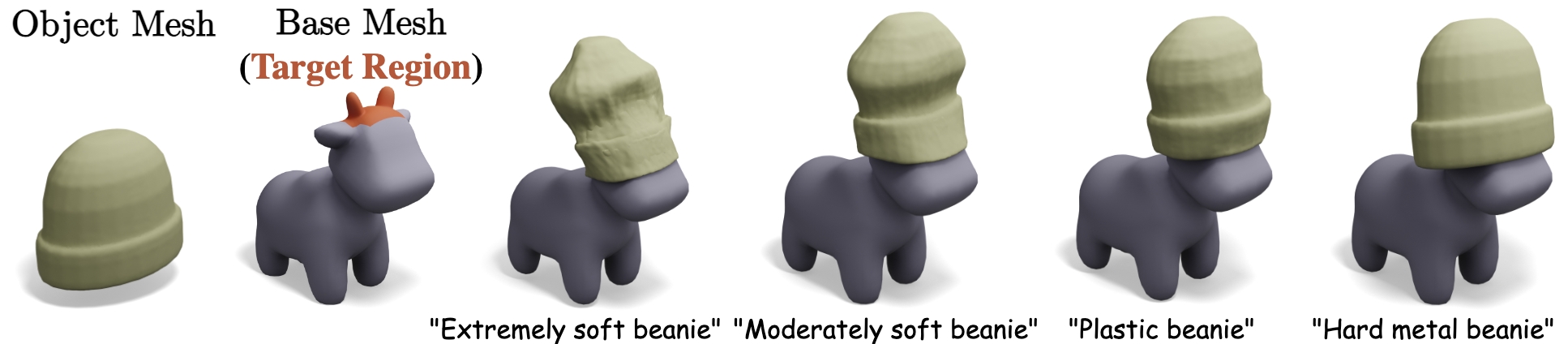}
    \captionof{figure}{\ourmethod{} is capable of fitting the accessory onto the base mesh in a way that considers material properties. Instead of having to specify the numeric material parameters, artists can provide material guidance through text prompts, that get combined with rendered images and interpreted by a VLM to output specific elastic parameters.}
    \label{fig:materials}
\end{figure*}%

\subsection{Step 3: Rigid finetuning}
\label{sec:step3}
The output of the previous section is a (scaled) rigid transformation $\rigid_2$ that moves the accessory reasonably close to the highlighted region of the base mesh without intersecting it; however, it likely fits too loosely to seem realistic (see \autoref{fig:main_method}, middle).

To refine this transformation, we draw inspiration from works in the physically-based simulation area, in which the relative positions of many objects are often assumed to minimize some physically meaningful energy (e.g., gravity) subject to non-intersecting constraints.
A common strategy for imposing these constraints is by adding repulsive energy terms to the optimization \cite{baraff1998large,li2020incremental,lan2024efficient}.
In a similar way, starting from $\rigid_2 = \rigid(\parameters_2)$, we will optimize the loss
\begin{equation}\label{equ:total-loss}
\mathcal{L}(\parameters) = \mathcal{L}_{p}(\parameters) + \mathcal{L}_{\text{IPC}}(\parameters)\,,
\end{equation}
where $\mathcal{L}_{p}$ is the proximity loss introduced in \autoref{sec:step1} and $\mathcal{L}_{\text{IPC}}$ is the \emph{Incremental Potential Contact} barrier energy \cite{li2020incremental}, given by
\begin{equation*}
\mathcal{L}_{\text{IPC}}(\parameters) = \sum_{\vertex_i \in \objectmesh} b_{\contactthreshold}\left(d\left(\rigid(\parameters)\vertex_i,\basemesh\right)\right)\,,
\label{contact_loss}
\end{equation*}
where
\begin{equation*}
    b_h(x) =
    \begin{cases}
      -(x-h)^{2}\ln\left( \dfrac{x}{h} \right) & \text{if}\quad x < h \\
      0       & \text{otherwise.}
    \end{cases}
\end{equation*}
This energy penalizes the object's vertices as they approach the base surface, with a second-order continuous (\( C^2 \)) form at the boundary \( d\left(\rigid(\parameters)\vertex_i,\basemesh\right) = \contactthreshold \), and naturally vanishes as \( d\left(\rigid(\parameters)\vertex_i,\basemesh\right) \to \contactthreshold \) or beyond (in our experiments, $h=0.01$). In addition to its use as an optimization energy in this section, we also use $\mathcal{L}_{\text{IPC}}$ to classify if a configuration is or not intersecting in the previous sections.

Minimizing the total loss from \autoref{equ:total-loss} using gradient-based optimization, starting from $\rigid_2$, results in a novel (scaled) rigid transformation $\rigid_3 = \rigid(\parameters^\star)$ that fits the object tightly without intersections over the base mesh.

\begin{figure*}[t]
\vspace{3mm}
\centering
\includegraphics[width=\textwidth]{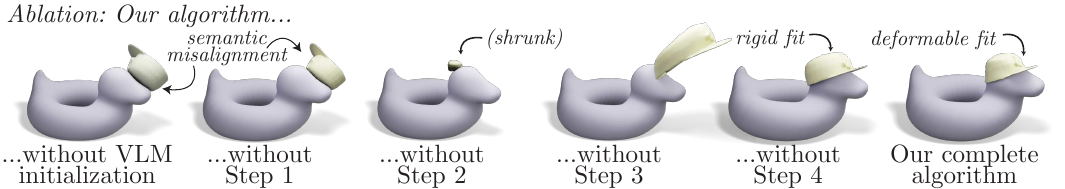}
\caption{\ourmethod{} uses a multi-step process, where each of them is critical for achieving a desirable mesh to mesh composition result. Omitting any of the steps yields a semantically (two left results) or physically (two center results) unplausible solutions. Utilizing Step 4 further improves the result (second from the right), allowing the asset to deform and better fit the base mesh (rightmost).} %
\label{fig:steps-ablation}
\end{figure*}

\subsection{Step 4: Elastic deformation}
\label{sec:step4}

The physically-inspired optimization of the previous step outputs a rigidly transformed mesh $\restmesh = \rigid_3 (\objectmesh) = ( \rigid_3 (\objectverts), \objectfaces)$, which we interpret as the best possible \emph{rigid fit} of the object onto the base mesh.
In reality, however, many accessories fit their wearer more tightly by undergoing small deformations: a hat is stretched by a head's shape, a necklace drapes over a person's chest, a face mask deforms to fit the nose and mouth of a surgeon.
Beyond physical deformation, an artist may also wish to perform small physical non-rigid edits on a general-purpose object mesh to better match their specific creative intentions.

To model this final fit, we will follow the lead of prior work \cite{aigerman2022neural, kim2025meshup, gao2023textdeformer, dinh2025geometry} and parametrize the object mesh deformation via a per-face Jacobian field \( \{J_i \in \mathbb{R}^{3 \times 3}\} \), which describes the local transformation of the surface. From these Jacobians, one can recover vertex-wise deformations $ \vertext_i = \transformation (\vertex_i) $  through a Poisson-like least-squares minimization:
\begin{equation}\label{equ:poisson}
\transformation^\star = \arg\min_t \sum_i a_i \left\| \nabla_i(t) - J_i \right\|_2^2\, ,
\end{equation}
where $a_i$ is the area of the $i$-th triangle in the mesh. This allows us to write $\transformation$ as a function of the per-face Jacobians $\transformation = \transformation(J_1, J_2, \dots) = \transformation(\{J_i\})$.
Our proximity and IPC losses from the previous section can thus be analogously defined for this non-rigid setting as
\begin{align}
\mathcal{L}_{p}(\{J_i\}) & = \sum_{\vertex_{i} \in \objectmesh} \hat{d} (\transformation(\{J_i\})\vertex_i, \highlightedmesh)^2  
+  \sum_{\vertex_{j} \in \highlightedmesh} \hat{d}(\vertex_j, \transformation(\{J_i\})\objectmesh)^2\,
\end{align}
and
\begin{align}
\mathcal{L}_{\text{IPC}}(\{J_i\}) & = \sum_{\vertex_i \in \objectmesh} b_{\contactthreshold}\left(d\left(\transformation(\{J_i\})\vertex_i,\basemesh\right)\right)\,,
\end{align}
where we use a lower proximity threshold value of $\ell=0.01$ to zero out distance entries, enforcing the deformation to be conservatively confined to its local boundries. 
The point-to-mesh distances $d$ are once again computed using a bounding volume hierarchy; however, this hierarchy is recomputed every $20$ iterations due to the presence of deformations.
Additionally, we follow the strategy proposed by \emph{MeshUp} \cite{kim2025meshup} and guide the deformation towards a semantically meaningful fit via Score Distillation Sampling (SDS).
Let \( z \) denote a rendered image of the current mesh, and \( \mathcal{T} \) a guiding text prompt. At each iteration, we sample a timestep \( t \sim \mathcal{U}(0,1) \) and a noise vector \( \epsilon \sim \mathcal{N}(0,I) \), and use the gradient
$$
\nabla_{J_i} \mathcal{L}_{\text{SDS}}(\{J_i\}) = w(t)\left(
\epsilon_\omega(z_t, \transformation(\{J_i\}), t) - \epsilon
\right)
\cdot
\frac{\partial z_t}{\partial J_i}.
$$
Finally, some deformations are more physically realistic than others, and different materials will deform by different amounts. To model this, we include the standard elastic NeoHookean energy as a loss term (see \cite{sifakis2012fem}, Chap. 3.6),
\begin{align*}
\mathcal{L}_e (\{J_i\}) = & \sum_i \left(\frac{\mu}{2}\left(\tr(J_i^\top J_i) - 3\right)\right.
- \left.\mu \log(\det(J_i))
 + \frac{\lambda}{2} \log^2(\det(J_i)) \right)
\end{align*}
where the material parameters $\lambda$ and $\mu$ are either given as input to our algorithm or inferred either from rendered images or text descriptions by our multi-agent initialization framework (see \autoref{sec:initialization}).
We end by minimizing the total loss
\[
\mathcal{L} = \mathcal{L}_{p} + \mathcal{L}_{\text{IPC}} + \lambda_{\text{SDS}} \mathcal{L}_{\text{SDS}} + \lambda_{e} \mathcal{L}_e\, ,
\]
with respect to $\{J_i\}$ using gradient-based optimization, and solve \autoref{equ:poisson} to obtain our final transformation $\transformation$ (see \autoref{fig:main_method}, right). As desired, $\transformation$ fits the object mesh tightly onto the base mesh in a semantically and physically realistic way without intersections, ending our algorithm.

\subsection{Initialization}
\label{sec:initialization}
\begin{figure*}[t]
    \vspace{3mm}
    \includegraphics[width=\textwidth]{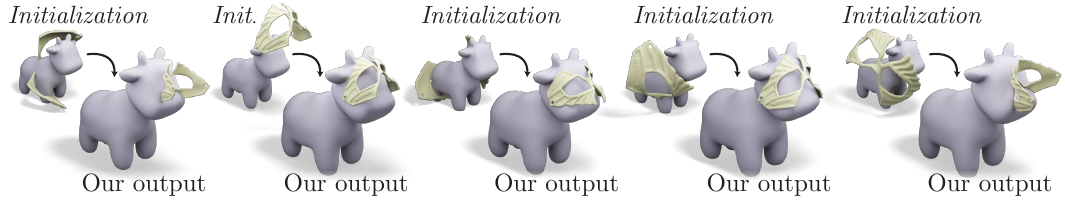}   
    \caption{Our method is moderately robust to different initialization configurations, although it can fail to converge to the desired output in very adversarial cases (leftmost and rightmost). In \autoref{sec:initialization}, we introduce a VLM initialization strategy that avoids these cases.}
    \label{fig:initialization}
    \vspace{-4mm}
\end{figure*}

Like any gradient-based optimization, our algorithm is sensitive to initialization; in particular, it depends on the initial relative configuration of the two meshes that is then modified in steps one through four.
Experimentally, we find that our method is relatively robust to different initial configurations; however, the most adversarial of them can cause it to converge to undesirable local minima (see \autoref{fig:initialization}).

To avoid these, we initialize the relative configuration using a VLM.  First, from a text description (e.g., "a hat") and renderings of the object, the VLM is prompted to produce a canonical reference frame for it (e.g., "the front of the hat is the region with a visor, the top part of the hat...."). This process is repeated for the base mesh. Then, several different random orientations are sampled for the object mesh. For each, a VLM is used to deduce its alignment with the canonical reference frame and with the base mesh's. The latter is used to score each configuration, and we select the highest-scoring one as initialization (see Supplementary Material for details).

 An additional agent uses user input text to output the the material parameters $\lambda, \mu$, giving users a fully-integrated, text-based control over the entire optimization process (see \autoref{fig:main_method}).
More details about our initialization scheme are provided in the Supplementary.

\section{Experiments} \label{sec:experiments}

Our algorithm is implemented in Python using \textsc{PyTorch} for autodifferentiation. Our comparisons to Iterative Closest Point and its derivative works, Deep Closest Point \cite{wang2019deep}, and Instant3dit \cite{barda2025instant3dit} use official or publicly available implementations \cite{Zhou2018}. We render our results using \textsc{Blender}, and provide more implementation details in the Supplementary Materials.

As is theoretically expected, our main computational bottleneck is the calculation of point-to-mesh distances involved in the proximity and IPC losses $\mathcal{L}_p$ and $\mathcal{L}_{IPC}$.
Our tailor-made BVH reduces the computational cost of this step, allowing us to produce results for relatively large meshes within reasonable times (see Supplementary).

\begin{wrapfigure}[10]{r}{0.50\linewidth}
    \vspace{-4mm}
    \centering
    \includegraphics[width=\linewidth]{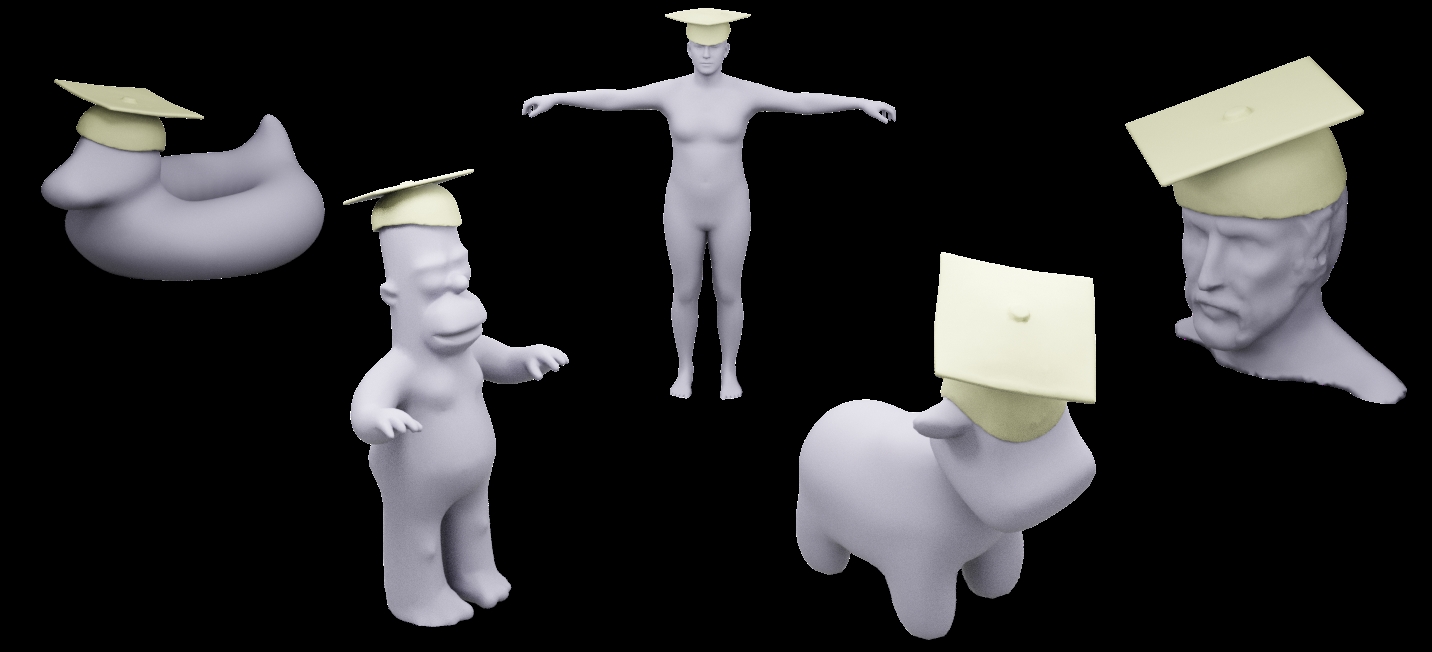}
    \caption{}
    \label{fig:same_accesory_different_base}
\end{wrapfigure}

A particularly relevant set of parameters are the material elastic properties $\lambda$ and $\mu$, which can be derived from the material's Young's modulus and Poisson's ratio \cite{sifakis2012fem}.
These quantities are tabulated for well-known materials; however, we make the process more intuitive for an artist by assigning the values from text inputs using an LLM. We explore both the VLM assignment and the effect of these parameters in \autoref{fig:materials}, in which the object is allowed to deform more or less depending on the specified material.

As expected, our algorithm's output is most sensitive to the selected target region on the base mesh $\highlightedfaces\subset \basefaces$.
In \autoref{fig:highlight-region}, we show how this input can allow artists to select different semantically meaningful assembly configurations between the two objects in ambiguous cases.
A similar example is given in \autoref{fig:four_fingers}, where different highlight regions specify the finger wearing a ring.

To obtain a tight, non-intersecting fit between the base and the accessory, our algorithm relies on a VLM-driven initialization and four successive optimizations, the input and output of each we show in \autoref{fig:main_method}. In \autoref{fig:steps-ablation}, we show the importance of each of these steps: skipping the VLM initialization results in a non-semantic fit (a backwards hat), while skipping any of the intermediate steps results in outputs with significantly worse physical and geometric fit. As expected, skipping Step 4 results in a strictly rigid fit, while enabling Step 4 allows the object to deform.

Critically, in Step 2 of our algorithm (\autoref{sec:step2}), we sample trajectories to find a close, non-intersecting configuration. We exemplify this in \autoref{fig:trajectory_method}; as shown in \autoref{fig:steps-ablation}, skipping this causes the barrier term in Step 3 to greatly displace the shape away from the base mesh.

\section{Comparisons} \label{sec:comparisons}
\subsection{Qualitative Evaluation}
Inspired by the digital artists' workflows, we consider the specific question of how to align two pre-existing shapes in a semantically meaningful, non-intersecting, tight way.
While the lack of prior work dedicated to this particular problem makes direct comparisons difficult, we justify the need for our method by showing the failure of existing general-purpose algorithms when applied to this task.

For example, by finding an alignment between two 3D objects, our method can be seen as a special kind of \emph{registration} algorithm. However, unlike standard registration algorithms, we do not rely on the existence of a rigid geometric match between the two objects: as shown in \autoref{fig:materials}, the accessory may need to deform significantly for it to even be a partial match of the base mesh.

\begin{figure}[t]
    \vspace{3mm}
    \centering
    
    \begin{subfigure}[t]{0.49\linewidth}
        \centering
        \begin{minipage}[t][0.8cm][t]{\linewidth}
            \centering\textbf{(a) Ring Placement}
        \end{minipage}
        \phantomsubcaption\label{fig:four_fingers}
        \includegraphics[width=\linewidth]{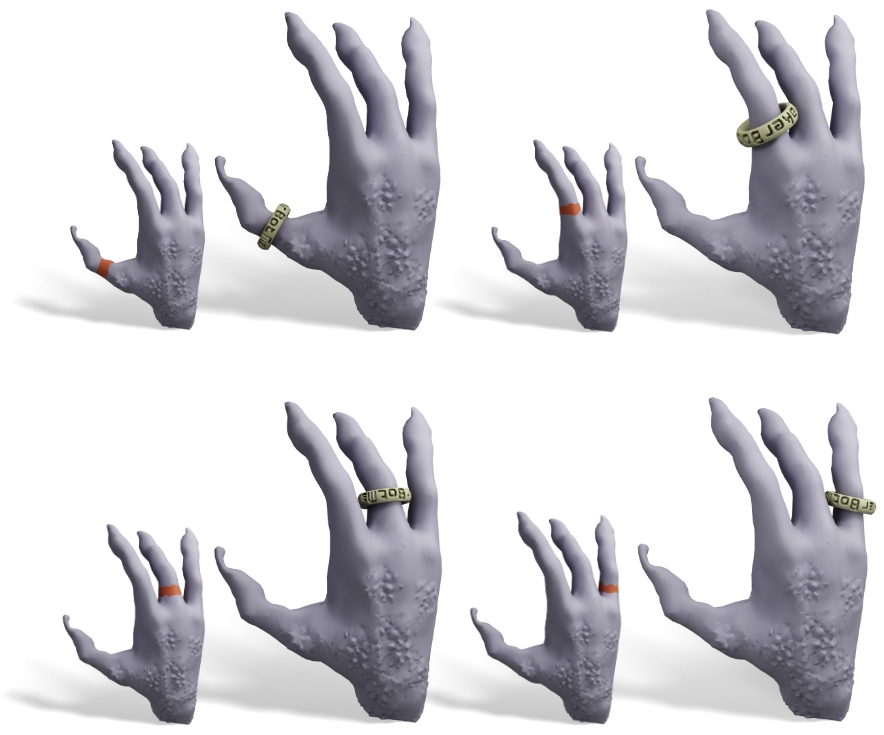}
    \end{subfigure}
    \hfill
    \begin{subfigure}[t]{0.49\linewidth}
        \centering
        \begin{minipage}[t][0.8cm][t]{\linewidth}
            \centering\textbf{(b) Region Sensitivity}
        \end{minipage}
        \phantomsubcaption\label{fig:highlight-region}
        \includegraphics[width=\linewidth]{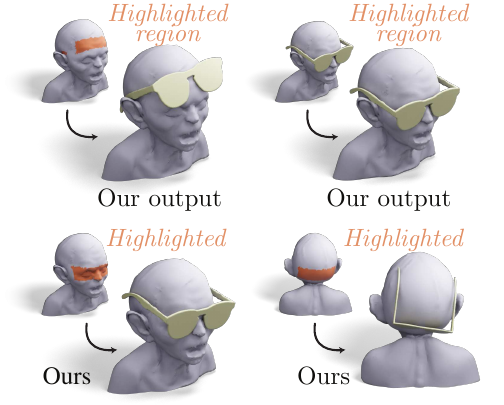}
    \end{subfigure}

    \vspace{3mm}
    \caption{
    \textbf{Region-controlled composition.}
    \textbf{(a)} Given four distinct user-selected target regions, \ourmethod{} places a ring precisely on each selected region, demonstrating explicit user control over the fitting process.
    \textbf{(b)} Because the method adheres strictly to the selected region, asset placement is sensitive to the region definition; for example, glasses sit naturally on the ears only if a small portion of the ear region is included in the selection.
    }
    \label{fig:region_control}
\end{figure}

\begin{figure*}[t]   
    \vspace{3mm}
    \centering
    \includegraphics[width=\textwidth]{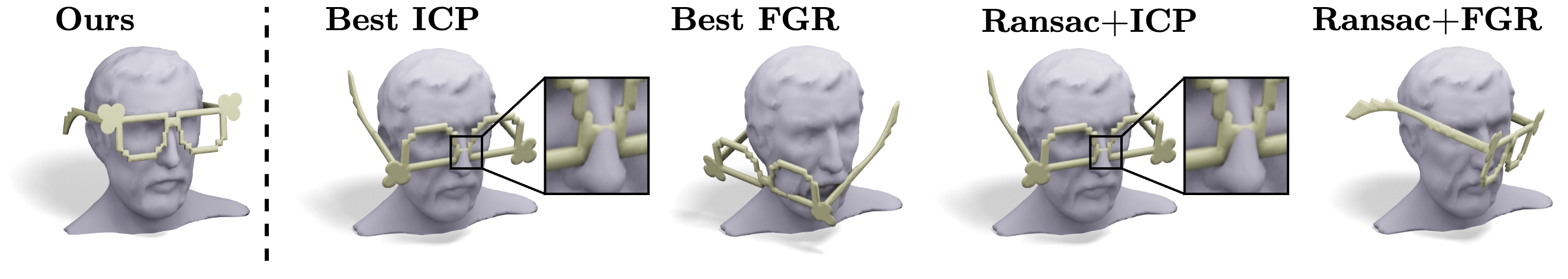}
        \captionof{figure}{Unlike ours (left), classical registration algorithms like ICP contain no semantic guidance: therefore, they will produce unrealistic configurations; e.g., glasses that sit on the eyes upside down. These algorithms are also not designed to avoid intersections; therefore, they will produce many of them (see blowups). See supplementary materials for details of these methods.}
    \label{fig:comparison}
\end{figure*}

More importantly, general shape registration algorithms, which are often used with point cloud inputs, will cause significant intersections between the base mesh and the object.
We show examples of this phenomenon in \autoref{fig:comparison}, where we compare our method to variants of classical registration methods \cite{besl1992method}\footnote{\label{fn:alignment_baseline}Classical registration methods are highly sensitive to bad initialization, so we use their variants that are known to be more robust. More details can be found in the Supplementary Materials.}.
Additionally, non-neural methods like those based on the Iterative Closest Point algorithm will not account for the semantic fit between the two objects; \eg, fitting the glasses backwards on the face.
We also experimented with public implementations of data-driven methods like Deep Closest Point \cite{wang2019dcp}; however, we found that these struggle greatly with most standard shapes outside of their training data, when restricted to the highlighted target region.

By modifying a shape using diffusion guidance, a highlighted region, and an optional text prompt, our work may seem similar to existing text-guided generative mesh-editing works.
Unlike these, however, our work preserves both the object and base meshes as well as any information stored in them.
Taking Instant3dit \cite{barda2025instant3dit} as a representative example, we show the difference between ours and this class of methods in \autoref{fig:instant3dit_comparison}.
Instant3dit produces a generative edit that merges the geometry of both objects and discards the animation rig of the base mesh and the parametrization of the accessory.
By contrast, our method places a world of possibilities in the hands of the artist by preserving all these, enabling common downstream tasks like animation, physical simulation, and texture painting.

\subsection{Quantitative Evaluation.}
We quantitatively evaluate our method along two complementary dimensions: semantic alignment quality and geometric validity. We also show a user study of semantic alignment in the Supplementary Materials.\\

In \autoref{tab:quant_all}, we  compare our method against multiple alignment baselines, including Instant3Dit~\cite{barda2025instant3dit}, Best ICP/FGR, and RANSAC-based variants~\ref{fn:alignment_baseline}.  
We measure semantic quality using VQA~\cite{lin2024evaluating} and two CLIP-based metrics~\cite{radford2021learning,wang2023exploring} on image renderings of ten examples. We post examples of these renderings in the Supplementary Materials.
Across these metrics, our full method achieves the strongest or near-strongest performance (for CLIP-IQA scores, the margin is $0.006\text{--}0.009$).
Alignment-based methods also frequently produce mesh intersections, an undesirable artifact for most graphics workflows. 
In \autoref{tab:penetration} we report total intersecting faces and maximum penetration depth for each for these methods, demonstrating that while all baseline methods exhibit substantial intersections, our method produces intersection-free compositions.

\section{Applications} \label{sec:applications}
Since our method is designed to fit directly into artists' workflows, we prioritize providing them with the highest degree of control over the fit, automating only the tedious rigid and deformable fitting procedure.
With our method, artists may specify every creative detail: from the input meshes and their properties, which are preserved (see \autoref{fig:instant3dit_comparison}), to the accessory's material parameters (see \autoref{fig:materials}) and the specific region of the fit (see \autoref{fig:four_fingers}).

\begin{table}[t]
\vspace{4mm}
\centering
\small
\setlength{\tabcolsep}{6pt}
\renewcommand{\arraystretch}{1.2}
\setlength{\aboverulesep}{0pt}
\setlength{\belowrulesep}{0pt}

\resizebox{\columnwidth}{!}{%
\begin{tabular}{lcccccc}
\toprule
\multicolumn{7}{l}{\textbf{Alignment Baselines}}\\
\rowcolor{headercolor}
Metric & \shortstack{Instant3Dit} & \shortstack{Best ICP} & \shortstack{Best FGR} & \shortstack{RANSAC+ICP} & \shortstack{RANSAC+FGR} & \shortstack{Ours} \\
\midrule
\rowcolor{zebra}
CLIP ($\uparrow$)     & 0.311 & 0.341 & 0.333 & 0.331 & 0.321 & \textbf{0.356} \\
CLIP-IQA ($\uparrow$) & 0.559 & 0.586 & 0.609 & 0.608 & \textbf{0.610} & 0.604 \\
\rowcolor{zebra}
VQA ($\uparrow$)      & 65.62 & 73.71 & 73.97 & 73.72 & 73.68 & \textbf{74.49} \\
\midrule
\\[4pt] 
\midrule
\multicolumn{7}{l}{\textbf{Ablation Study}}\\
\rowcolor{headercolor}
Metric & \shortstack{Init} & \shortstack{Step 1} & \shortstack{Step 2} & \shortstack{Step 3} & \shortstack{Step 4} & \shortstack{Ours} \\
\midrule
\rowcolor{zebra}
CLIP ($\uparrow$)     & 0.338 & 0.351 & 0.346 & 0.347 & 0.354 & \textbf{0.356} \\
CLIP-IQA ($\uparrow$) & 0.594 & 0.605 & 0.609 & 0.595 & \textbf{0.613} & 0.604 \\
\rowcolor{zebra}
VQA ($\uparrow$)      & 74.197 & 73.967 & 73.776 & 74.049 & 73.985 & \textbf{74.49} \\
\bottomrule
\end{tabular}%
}
\vspace{4mm}
\caption{Quantitative evaluation using CLIP, CLIP-IQA, and VQA scores. Top: comparison against alignment baselines. Bottom: ablation study of our multi-step pipeline. Scores are averaged over per-example best renderings (best of 4); higher is better.}
\vspace{-2mm}
\label{tab:quant_all}
\end{table}

We exemplify our algorithm's applicability through several prototypical examples. As shown in \autoref{fig:gallery}, our method can find tight, non-intersecting configurations between a diverse range of meshes, even when these contain complex geometries, concavities, and topologies.
In \autoref{fig:teaser}, successive pre-existing shapes are added to an existing model to create a superhero character.
Conversely, in \autoref{fig:same_accesory_different_base}, the same accessory is automatically placed on a wide range of characters, demonstrating how our method can adapt to different scales, head shapes, and body proportions.

\begin{table}[t]
\vspace{4mm}
\centering
\small
\setlength{\tabcolsep}{3pt}
\renewcommand{\arraystretch}{1.15}

\begin{tabularx}{\linewidth}{>{\raggedright\arraybackslash}X c c c c c}
\toprule
{\textbf{Ablation Study}}\\
\rowcolor{headercolor}
Metric &
Best ICP &
Best FGR &
RANSAC+ICP &
RANSAC+FGR &
Ours \\
\midrule
\rowcolor{zebra}
Intersecting faces ($\downarrow$)
& 628 & 660 & 544 & 434 & \textbf{0} \\
Max penetration ($\downarrow$)
& 0.0197 & 0.0237 & 0.0243 & 0.0237 & \textbf{0} \\
\bottomrule
\end{tabularx}
\vspace{4mm}
\caption{Penetration statistics averaged over ten experiments. We report the total number of intersecting faces and the maximum penetration depth. Our method produces intersection-free compositions (all zeros).}
\vspace{-3mm}
\label{tab:penetration}
\end{table}

\section{Conclusion} \label{sec:conclusion}
We introduced a mesh composition algorithm that produces a meaningful, tight, non-intersecting fit between two shapes, inspired by the application of 3D digital art.
For our work to be useful in an interactive, real-time 3D modeling session, however, its runtime would need to improve significantly.
We believe there are several potential avenues for this: for example, one could dramatically improve runtimes by building increasingly fine nested simulation cages using existing methods \cite{sacht2015nested}. Then, one could perform the optimizations in our algorithm successively for each refinement level, thus reducing the need for extensive computation on the finest of levels.

As a tool of artist control, our algorithm requires a user to specify a highlighted region of the base object onto which the accessory should be attached. Experimentally, we find that our algorithm's output is sensitive to small changes in this region (see \autoref{fig:highlight-region}), which could lead to user frustration.
This issue could be resolved by future work combining our algorithm with existing text-based mesh selection tools like \emph{3D Highlighter} \cite{decatur20233d}.

As a final step in our fit, our algorithm allows the object to deform in order to better match the base mesh.
This results in an improved fit (see \autoref{fig:steps-ablation}) and provides an additional avenue for artist control (see \autoref{fig:materials}).
However, our method still assumes mostly-rigid scenarios in which the attachment to the base mesh is the main source of deformation.
For fundamentally elastic materials in which other internal and external forces are dominant (\eg, a piece of cloth draping over a character), a better fit can likely be achieved with existing physical simulation tools.

This is a time of friction between rapidly improving 3D AI research and artist communities reluctant to incorporate these tools due to loss of control and creativity.
While solving this friction is a complex problem beyond the scope of any one paper, we hope that non-generative AI-driven tools like \ourmethod{}, which is intended to fit directly into real digital artists' workflows while maximizing creative control, can aid in building bridges between these two worlds.

\newpage
{\small
\bibliographystyle{splncs04}
\bibliography{references}

\begin{thebibliography}{10}
\providecommand{\url}[1]{\texttt{#1}}
\providecommand{\urlprefix}{URL }
\providecommand{\doi}[1]{https://doi.org/#1}

\bibitem{aigerman2022neural}
Aigerman, N., Gupta, K., Kim, V.G., Chaudhuri, S., Saito, J., Groueix, T.: Neural jacobian fields: Learning intrinsic mappings of arbitrary meshes. arXiv preprint arXiv:2205.02904  (2022)

\bibitem{aoki2019pointnetlk}
Aoki, Y., Goforth, H., Srivatsan, R.A., Lucey, S.: Pointnetlk: Robust \& efficient point cloud registration using pointnet. In: Proceedings of the IEEE/CVF Conference on Computer Vision and Pattern Recognition. pp. 7163--7172 (2019)

\bibitem{baraff1998large}
Baraff, D., Witkin, A.: Large steps in cloth simulation. In: Proceedings of the 25th annual conference on Computer graphics and interactive techniques. pp. 43--54 (1998)

\bibitem{barda2025instant3dit}
Barda, A., Gadelha, M., Kim, V.G., Aigerman, N., Bermano, A.H., Groueix, T.: Instant3dit: Multiview inpainting for fast editing of 3d objects. In: Proceedings of the Computer Vision and Pattern Recognition Conference. pp. 16273--16282 (2025)

\bibitem{besl1992method}
Besl, P.J., McKay, N.D.: Method for registration of 3-d shapes. In: Sensor fusion IV: control paradigms and data structures. vol.~1611, pp. 586--606. Spie (1992)

\bibitem{botschsorkine08deformsurvey}
Botsch, M., Sorkine, O.: On linear variational surface deformation methods. IEEE Transactions on Visualization and Computer Graphics  \textbf{14}(1),  213--230 (2008). \doi{10.1109/TVCG.2007.1054}

\bibitem{chen2024textmeshrefine}
Chen, Y.C., Ling, S., Chen, Z., Kim, V.G., Gadelha, M., Jacobson, A.: Text-guided controllable mesh refinement for interactive 3d modeling. In: ACM SIGGRAPH Asia (2024)

\bibitem{choy2020dgr}
Choy, C.B., Dong, W., Koltun, V.: Deep global registration. In: Proceedings of the IEEE/CVF Conference on Computer Vision and Pattern Recognition (CVPR). pp. 2514--2523 (2020)

\bibitem{decatur20233d}
Decatur, D., Lang, I., Hanocka, R.: 3d highlighter: Localizing regions on 3d shapes via text descriptions. In: Proceedings of the IEEE/CVF Conference on Computer Vision and Pattern Recognition. pp. 20930--20939 (2023)

\bibitem{dinh2025geometry}
Dinh, N.A., Lang, I., Kim, H., Stein, O., Hanocka, R.: Geometry in style: 3d stylization via surface normal deformation. In: Proceedings of the Computer Vision and Pattern Recognition Conference (CVPR). pp. 28456--28467 (2025)

\bibitem{Fulton:LSD:2018}
Fulton, L., Modi, V., Duvenaud, D., Levin, D.I.W., Jacobson, A.: Latent-space dynamics for reduced deformable simulation. Computer Graphics Forum  (2019)

\bibitem{gao2023textdeformer}
Gao, W., Aigerman, N., Groueix, T., Kim, V.G., Hanocka, R.: Textdeformer: Geometry manipulation using text guidance. In: ACM Transactions on Graphics (SIGGRAPH) (2023)

\bibitem{guo2025craft}
Guo, M., Tang, M., Cha, H., Zhang, R., Liu, C.K., Wu, J.: Craft: Designing creative and functional 3d objects. In: 2025 IEEE/CVF Winter Conference on Applications of Computer Vision (WACV). pp. 7215--7224. IEEE (2025)

\bibitem{hao2024meshtron}
Hao, Z., Romero, D.W., Lin, T.Y., Liu, M.Y.: Meshtron: High-fidelity, artist-like 3d mesh generation at scale. arXiv preprint arXiv:2412.09548  (2024)

\bibitem{huang2020fmr}
Huang, S., Liang, Z., Cho, Y., Li, X., Wang, Y., Yang, Y.: Feature-metric registration: A fast, robust, feature-metric deep learning method for point cloud registration. In: Proceedings of the IEEE/CVF Conference on Computer Vision and Pattern Recognition (CVPR). pp. 9969--9978 (2020)

\bibitem{skinningcourse:2014}
Jacobson, A., Deng, Z., Kavan, L., Lewis, J.: Skinning: Real-time shape deformation. In: ACM SIGGRAPH 2014 Courses (2014)

\bibitem{jiang2023se}
Jiang, H., Salzmann, M., Dang, Z., Xie, J., Yang, J.: Se (3) diffusion model-based point cloud registration for robust 6d object pose estimation. In: Thirty-seventh Conference on Neural Information Processing Systems (2023)

\bibitem{kerbl20233d}
Kerbl, B., Kopanas, G., Leimk{\"u}hler, T., Drettakis, G.: 3d gaussian splatting for real-time radiance field rendering. ACM Trans. Graph.  \textbf{42}(4),  139--1 (2023)

\bibitem{kim2025meshup}
Kim, H., Lang, I., Aigerman, N., Groueix, T., Kim, V.G., Hanocka, R.: Meshup: Multi-target mesh deformation via blended score distillation. In: 2025 International Conference on 3D Vision (3DV). pp. 222--239. IEEE (2025)

\bibitem{lan2024efficient}
Lan, L., Lu, Z., Long, J., Yuan, C., Li, X., He, X., Wang, H., Jiang, C., Yang, Y.: Efficient gpu cloth simulation with non-distance barriers and subspace reuse. arXiv preprint arXiv:2403.19272  (2024)

\bibitem{lang2021dpc}
Lang, I., Ginzburg, D., Avidan, S., Raviv, D.: {DPC: Unsupervised Deep Point Correspondence via Cross and Self Construction}. In: Proceedings of the International Conference on 3D Vision (3DV). pp. 1442--1451 (2021)

\bibitem{li2020incremental}
Li, M., Ferguson, Z., Schneider, T., Langlois, T.R., Zorin, D., Panozzo, D., Jiang, C., Kaufman, D.M.: Incremental potential contact: intersection-and inversion-free, large-deformation dynamics. ACM Trans. Graph.  \textbf{39}(4), ~49 (2020)

\bibitem{lin2023magic3d}
Lin, C.H., Gao, J., Tang, L., Takikawa, T., Zeng, X., Huang, X., Kreis, K., Fidler, S., Liu, M.Y., Lin, T.Y.: Magic3d: High-resolution text-to-3d content creation. In: Proceedings of the IEEE/CVF conference on computer vision and pattern recognition. pp. 300--309 (2023)

\bibitem{lin2024evaluating}
Lin, Z., Pathak, D., Li, B., Li, J., Xia, X., Neubig, G., Zhang, P., Ramanan, D.: Evaluating text-to-visual generation with image-to-text generation. In: European Conference on Computer Vision. pp. 366--384. Springer (2024)

\bibitem{litany2017deep}
Litany, O., Remez, T., Rodolà, E., Bronstein, A.M., Bronstein, M.M.: Deep functional maps: Structured prediction for dense shape correspondence. In: Proceedings of the IEEE International Conference on Computer Vision (ICCV). pp. 5660--5668. IEEE Computer Society (Oct 2017). \doi{10.1109/ICCV.2017.603}, \url{https://openaccess.thecvf.com/content_ICCV_2017/papers/Litany_Deep_Functional_Maps_ICCV_2017_paper.pdf}

\bibitem{mckay2003review}
McKay, N.D.: 3d registration: A review of techniques. In: Proceedings of the SPIE Videometrics VIII (2003)

\bibitem{michel2022text2mesh}
Michel, O., Bar-On, R., Liu, R., Benaim, S., Hanocka, R.: Text2mesh: Text-driven neural stylization for meshes. In: Proceedings of the IEEE/CVF conference on computer vision and pattern recognition (CVPR). pp. 13492--13502 (2022)

\bibitem{mildenhall2021nerf}
Mildenhall, B., Srinivasan, P.P., Tancik, M., Barron, J.T., Ramamoorthi, R., Ng, R.: Nerf: Representing scenes as neural radiance fields for view synthesis. Communications of the ACM  \textbf{65}(1),  99--106 (2021)

\bibitem{radford2021learning}
Radford, A., Kim, J.W., Hallacy, C., Ramesh, A., Goh, G., Agarwal, S., Sastry, G., Askell, A., Mishkin, P., Clark, J., et~al.: Learning transferable visual models from natural language supervision. In: International conference on machine learning. pp. 8748--8763. PmLR (2021)

\bibitem{sacht2015nested}
Sacht, L., Vouga, E., Jacobson, A.: Nested cages. ACM Transactions on Graphics (TOG)  \textbf{34}(6),  1--14 (2015)

\bibitem{salvi2007survey}
Salvi, J., Matabosch, C., Fofi, D., Forest, J.: A review of recent registration methods for 3d modelling. Image and Vision Computing  \textbf{25}(5),  578--596 (2007)

\bibitem{segal2009generalized}
Segal, A., Haehnel, D., Thrun, S.: Generalized-icp. In: Robotics: science and systems. vol.~2, p.~435. Seattle, WA (2009)

\bibitem{shirley2009fundamentals}
Shirley, P., Ashikhmin, M., Marschner, S.: Fundamentals of computer graphics. AK Peters/CRC Press (2009)

\bibitem{sifakis2012fem}
Sifakis, E., Barbic, J.: Fem simulation of 3d deformable solids: a practitioner's guide to theory, discretization and model reduction. In: Acm siggraph 2012 courses, pp. 1--50 (2012)

\bibitem{sorkine2007rigid}
Sorkine, O., Alexa, M.: As-rigid-as-possible surface modeling. In: Symposium on Geometry processing. vol.~4, pp. 109--116. Citeseer (2007)

\bibitem{sorkine2004laplacian}
Sorkine, O., Cohen-Or, D., Lipman, Y., Alexa, M., R{\"o}ssl, C., Seidel, H.P.: Laplacian surface editing. In: Proceedings of the 2004 Eurographics/ACM SIGGRAPH symposium on Geometry processing. pp. 175--184 (2004)

\bibitem{tam2013registration}
Tam, G.K.L., Cheng, K., Lai, Y.K., Langbein, F.C., Liu, Y., Marshall, D., Martin, R.R., Sun, X.F., Rosin, P.L.: Registration of 3d point clouds: A survey. IEEE Transactions on Visualization and Computer Graphics  \textbf{19}(7),  1199--1217 (2013)

\bibitem{gemini}
Team, G., Anil, R., Borgeaud, S., Alayrac, J.B., Yu, J., Soricut, R., Schalkwyk, J., Dai, A.M., Hauth, A., Millican, K., et~al.: Gemini: a family of highly capable multimodal models. arXiv preprint arXiv:2312.11805  (2023)

\bibitem{wang2023exploring}
Wang, J., Chan, K.C., Loy, C.C.: Exploring clip for assessing the look and feel of images. In: Proceedings of the AAAI conference on artificial intelligence. vol.~37, pp. 2555--2563 (2023)

\bibitem{wang2019dcp}
Wang, Y., Solomon, J.M.: Deep closest point: Learning representations for point cloud registration. In: Proceedings of the IEEE/CVF International Conference on Computer Vision (ICCV) (October 2019)

\bibitem{wang2019deep}
Wang, Y., Solomon, J.M.: Deep closest point: Learning representations for point cloud registration. In: Proceedings of the IEEE/CVF international conference on computer vision. pp. 3523--3532 (2019)

\bibitem{xu2024fusiondeformer}
Xu, H., Wu, Y., Tang, X., Zhang, J., Zhang, Y., Zhang, Z., Li, C., Jin, X.: Fusiondeformer: Text-guided mesh deformation using diffusion models. The Visual Computer  \textbf{40}(7),  4701--4712 (2024)

\bibitem{xu2024instantmeshefficient3dmesh}
Xu, J., Cheng, W., Gao, Y., Wang, X., Gao, S., Shan, Y.: Instantmesh: Efficient 3d mesh generation from a single image with sparse-view large reconstruction models (2024), \url{https://arxiv.org/abs/2404.07191}

\bibitem{yang2024dreammesh}
Yang, H., Chen, Y., Pan, Y., Yao, T., Chen, Z., Wu, Z., Jiang, Y.G., Mei, T.: Dreammesh: Jointly manipulating and texturing triangle meshes for text-to-3d generation. In: European Conference on Computer Vision (ECCV) (2024)

\bibitem{yang2015go}
Yang, J., Li, H., Campbell, D., Jia, Y.: Go-icp: A globally optimal solution to 3d icp point-set registration. IEEE transactions on pattern analysis and machine intelligence  \textbf{38}(11),  2241--2254 (2015)

\bibitem{yew2022regtr}
Yew, Z.J., Lee, G.h.: Regtr: End-to-end point cloud correspondences with transformers. In: Proceedings of the IEEE/CVF Conference on Computer Vision and Pattern Recognition (CVPR) (2022)

\bibitem{yu2004poisson}
Yu, Y., Zhou, K., Xu, D., Shi, X., Bao, H., Guo, B., Shum, H.Y.: Mesh editing with poisson-based gradient field manipulation. In: ACM SIGGRAPH 2004 Papers. p. 644–651. SIGGRAPH '04, Association for Computing Machinery, New York, NY, USA (2004). \doi{10.1145/1186562.1015774}, \url{https://doi.org/10.1145/1186562.1015774}

\bibitem{zhou2016fast}
Zhou, Q.Y., Park, J., Koltun, V.: Fast global registration. In: European conference on computer vision. pp. 766--782. Springer (2016)

\bibitem{Zhou2018}
Zhou, Q.Y., Park, J., Koltun, V.: {Open3D}: {A} modern library for {3D} data processing. arXiv:1801.09847  (2018)

\bibitem{zhou2019continuity}
Zhou, Y., Barnes, C., Lu, J., Yang, J., Li, H.: On the continuity of rotation representations in neural networks. In: Proceedings of the IEEE/CVF conference on computer vision and pattern recognition. pp. 5745--5753 (2019)

\end{thebibliography}
}
\newpage

\appendix
\def\paperID{} %

\title{Supplementary Materials}
\author{}
\institute{}

\maketitle
\vspace{0.8em}

\section{Technical Details}
\label{sec:technical_details}
In this section, we will delineate some of the technical details and derivations we have left out in the method section of our main paper. 

\subsection{Distance Measures}
\label{subsec:distance_measures}

Recall that at the core of both our proximity loss $\mathcal{L}_p$ and the \textit{Incremental Potential Contact} barrier energy, $\mathcal{L}_{IPC}$ is the calculation of a vertex-to-surface distance measure between the two meshes. 

Let $\objectmesh$ denote our object mesh and $\basemesh$ our base mesh. The objective of this distance metric is to give an efficient measure of distance between the user-defined localized region on the base mesh, denoted by $\highlightedmesh$, and $\objectmesh$, and vice-versa. The minimum distance between the surface of any two triangular meshes is the minimum of the minimum pairwise edge-to-edge and triangle-to-vertex distance.
 In practice, we observe that using only the triangle-to-vertex distance gives us a practically ``good-enough" measure of the surface-wise distance without causing too much memory overhead. Intuitively, for measuring the vertex-to-triangle distance from $\objectmesh$ to $\highlightedmesh$, we calculate the distance from all vertices on $\objectmesh$, $v_i$, to the triangles on $\highlightedmesh$, $\basefaces$.
In order to reduce memory overhead in gradient back-propagation, we follow \cite{li2020incremental} and create a masked distance function $\hat{d}$ by thresholding the minimum triangle-to-vertex distance from each vertex $v_i$ on the object mesh to the triangles on the localized surface of the base mesh, $\highlightedmesh$:
\[
\hat{d}(v_i, \highlightedmesh) = 
\begin{cases}
d(v_i, \highlightedmesh) & \text{if } \min_{\mathcal{F}_b \in \highlightedmesh} d(v_i, \mathcal{F}_b) < \thresholdprox,\\
0 & \text{otherwise},
\end{cases}
\]
where $\thresholdprox = 0.5$. We discarding any triangle-to-point distance pairs that are far enough so their gradients have no meaningful influence over the alignment process. 

\paragraph{Proximity Loss.}
Using this measure, we approximate the proximity between the two mesh surfaces by calculating the vertex-to-face measure, $\hat{d}$, for both $\highlightedmesh$ and $\objectmesh$.

Our proximity loss is naturally defined as 
\begin{equation}
\mathcal{L}_{p}(\parameters) = \sum_{\vertex_{i} \in \objectmesh} \hat{d} (\rigid(\parameters)\vertex_i, \highlightedmesh)^2 +  \sum_{\vertex_{j} \in \highlightedmesh} \hat{d}(\vertex_j, \rigid (\parameters)\objectmesh)^2\, ,
\label{proximity_loss}
\end{equation}

During optimization, we update the pairwise triangle-to-vertex distance measure, $\min_{\mathcal{F}_b \in \highlightedmesh}$, thereby iteratively updating which triangle-vertex pairs are to be brought closer through the optimization process. Our key insight is that by iteratively updating these pairs, the object mesh $\objectmesh$ eventually moves in a direction that not only minimizes the shortest distance between the two surfaces, but also induces as many triangle-vertex pairs to be in close contact to each other, eventually finding a good ``fit" between the two meshes.

\begin{figure*}[t]
    \centering
    \vspace{-0.5cm}
    \includegraphics[width=\textwidth]{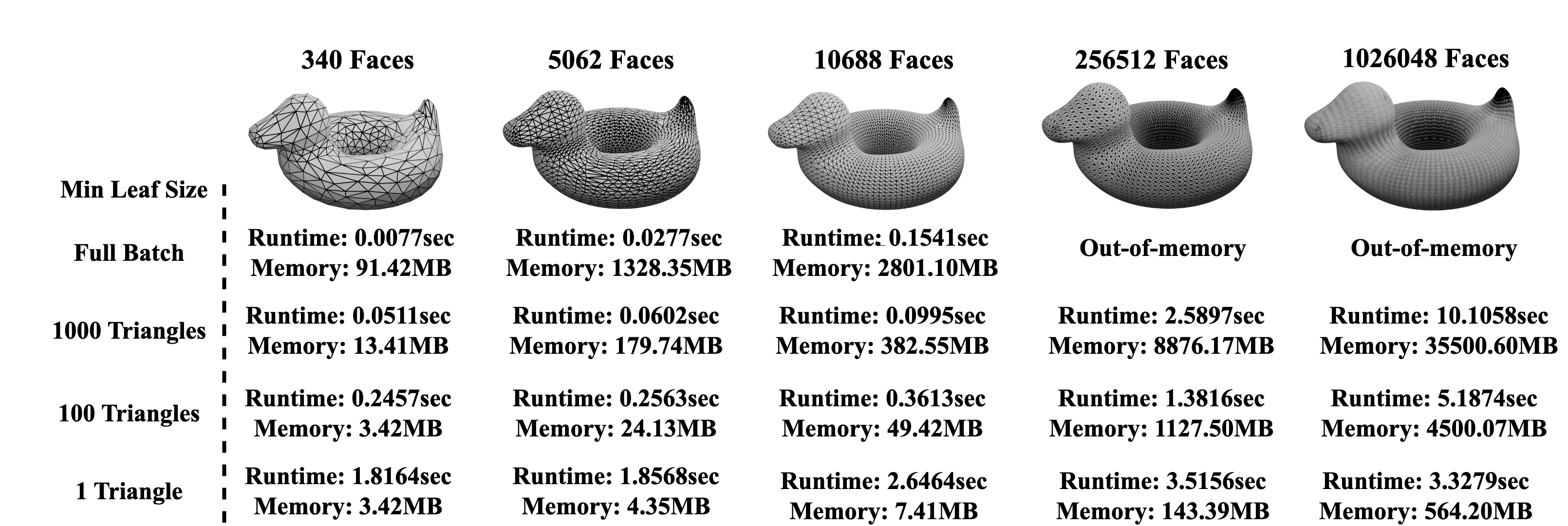}
    \vspace{-0.7cm}
    \captionof{figure}{We show a comparison of runtime and GPU memory consumption for calculating a single distance measure $d$ using our GPU-optimized BVH structure. We run the calculation with varying resolutions of the base mesh (visualized on top of each mesh), without using any user defined masks. We use a target mesh with a constant size of 8646 vertices and 17288 faces.}
    \label{fig:bvh_quant}
\end{figure*}%

\paragraph{Barrier Loss.}
We also reuse the contact measure $d(v_i, \mathcal{F}_b)$ to derive a loss term that prevents physical intersection between the two surfaces. 
We define a contact set based on a smaller threshold:
\[
\mathcal{C} = \{ i \mid \min_{\mathcal{F}_b \in \highlightedmesh} d(v_i, \mathcal{F}_b) < \ell \},
\]
where $\contactthreshold$ is a much smaller value than that of the proximity loss ($\ell=0.01$). This is because in order to prevent contact, we only need to account for the triangle-to-vertex pairs that are close enough such  to cause surface intersections.  

To differentiably penalize against proximity for these near-contact vertices, we adopt the clamped log-barrier energy from \cite{li2020incremental}. Given the pre-filtered distances $d_i$ corresponding to vertices in $\mathcal{C}$, the barrier energy is defined as:
\[
b_{\contactthreshold}(d_i) = 
-(d_i - \contactthreshold)^2 \ln\left( \dfrac{d_i}{\contactthreshold} \right),
\qquad \text{for } d_i < \contactthreshold.
\]
This energy smoothly penalizes points as they approach the base surface, with a second-order continuous (\( C^2 \)) form at the boundary $d_i = \contactthreshold$, and naturally vanishes as $d_i \to \contactthreshold$ or beyond. As a result, the total contact loss is expressed as:
\begin{equation}
\mathcal{L}_{\text{IPC}}(\parameters) = \sum_{\vertex_i \in \objectmesh} b_{\contactthreshold}\left(d\left(\rigid(\parameters)\vertex_i,\basemesh\right)\right)\,.
\label{contact_loss}
\end{equation}

We additionally note that this distance measure is implemented as a batched tensorized operation in Pytorch, saving us from for-looping through the entire list of vertices and faces.

\subsection{Vector-Optimized Bounding Volume Hierarchy}
\label{subsec:bvh}

Unfortunately, calculating $d(v_i, \mathcal{F}_b)$ requires $\mathcal{O}(|\objectverts||\basefaces|)$ computational complexity, where $\objectverts$ is the number of vertices and $\basefaces$ the number of faces involved. In other words, the computation could become infeasibly expensive as mesh resolution scales.

To overcome this challenge while adequately taking advantage of our method's tensorized nature, we employ a  Vector-Optimized Bounding Volume Hierarchy (BVH) structure that efficiently parses through a BVH tree before it reaches an optimal depth and falls back to a tensorized batch query of all triangle instances within its child nodes. 
\paragraph{Building the Tree}
To build the vector-optimized BVH over a face subset $\mathcal{F}$, our goal is to recursively calculate and store the minimum and maximum bounding box, $\mathbf{b}_{\min}=\min_{f\in[s:e)}\mathbf{m}_f$ and 
$\mathbf{b}_{\max}=\max_{f\in[s:e)}\mathbf{M}_f$
for each node, where $[s:e)$ is range of triangle indices in the node (and its children). Beginning with the top-most parent node with $[s:e)=[0,num(F))$ and ${(\mathbf{b}_{\min}, \mathbf{b}_{\max}})$ that encompass all triangles $\mathcal{F}$, we select the longest axis of this bounding box and sort the child nodes according to the center of their bounding boxes along the selected axis, ${c_i}$. We then split each node at the median of this sorted list, a process that we recursively repeat until the number of child node reaches a user-specified $\texttt{leaf\_size}$. The key contribution of our vector-optimized BVH is that we stop building this tree when $e-s\le\texttt{leaf\_size}$, where $\texttt{leaf\_size}$ is a variable set either by the user, or by some memory specification. To optimally take advantage of the vectorized nature of our distance calculation while preventing memory overhead from massive matrix multiplication, it is optimal to stop the tree traversal at a gpu-specific depth, and do a batched calculation of $d(n_{\text{vert}}^{(i)}, T)$ over triangles in the range ${f\in[s:e)}$. For simplicity, we set $\texttt{leaf\_size}=1000$ for all experiments run in this paper. 

\paragraph{Hybrid Best First Traversal}
The main challenge of using a BVH $\objectmesh$ is that the mesh is continuously transformed each iteration. For rigid transformations (rotation, translation, and scale), the BVH bounding boxes can be efficiently updated by transforming their centers and recalculating new, slightly looser half-extents.

For non-rigid transformations, we recalculate BVH structure every $cycle=20$ iterations, assuming the boxes stay relatively axis-aligned and the vertex positions relatively stable. This allows us to reuse the same BVH structure for small number of cycles with minimal sacrifice in accuracy.

\subsection{VLM}
\label{subsec:vlm}

Now we detail how we use Vision Language Models (VLM) to obtain a coarse initialization for Step 1.

Given randomly initialized meshes $\objectmesh$ and $\basemesh$, we aim for a good initialization of $\objectmesh$ by optimizing its orientation using a Vision-to-Language Model (VLM), while determining scale and translation heuristically. We utilize a multi-agent VLM pipeline responsible for defining semantic alignment criteria and classifying rendered views to output a compatibility score. 

The role of the first agent, the \textit{Describer} (Table \ref{tab:prompt_describer}), is to define a consistent semantic reference frame for both objects. Given the text labels of $\objectmesh$ and $\basemesh$ (and optionally rendered images), the agent constructs an \textbf{object-centric canonical coordinate frame} for each object. This frame specifies (i) an origin on the object (e.g., a natural landmark such as the center of the head or the midpoint between lenses), and (ii) three semantic axes: a rightward axis $(+X)$, an upward axis $(+Y)$, and a forward-facing axis $(+Z)$. 

Using these axes, the Describer defines the canonical views of each object—namely which parts of the object correspond to the \emph{front}, \emph{back}, \emph{left}, \emph{right}, \emph{top}, and \emph{bottom}. For example, for glasses the front view corresponds to the side where the lenses face outward, while the back view corresponds to the side facing the wearer. In addition to these axis definitions, the Describer also produces textual cues that help distinguish between ambiguous viewpoints (e.g., differences between left and right sides or between front and back). These semantic rules allow subsequent agents to interpret rendered images of the object in a consistent coordinate system regardless of the camera viewpoint.

In parallel, a second agent, the \textit{Material Verifier} (Table \ref{tab:prompt_material}), uses the material description generated by the Describer to infer approximate physical parameters of the target object, specifically Young's modulus and Poisson's ratio.

The second agent, the \textit{Scorer} (Table \ref{tab:prompt_scorer}), samples several candidate orientations by randomly rotating each object, and aligns $\objectmesh$ with $\basemesh$ by deducing their relative alignment from the candidate orientation that VLM deems to find most consistent (in other words, from whatever candidate orientation that VLM gets least confused by). 
Given each candidate orientation, we render 4 uniformly-sampled views of both objects. For each rendered image, the scorer predicts two simple properties. First, it determines which side of the object (more precisely, which canonical reference frame) is facing the camera,
\[
\texttt{visible\_face} \in 
\{\texttt{front},\texttt{back},\texttt{left},\texttt{right},\texttt{top},\texttt{bottom},\texttt{other}\}.
\]
Second, it determines how the object's upward direction appears in the image,
\[
\texttt{projected\_up} \in 
\{\texttt{up},\texttt{down},\texttt{left},\texttt{right},\texttt{other}\}.
\]
Intuitively, the first label describes which side of the object (or which side of the canonical frame) the camera sees, while the second label indicates whether the object appears upright, upside down, or rotated sideways in the image.

Before evaluating candidate orientations of $\objectmesh$, we first establish a consistent reference orientation for $\basemesh$. We do this by sampling several uniform azimuthal rotations of $\basemesh$, and selecting the rotation that produces the most self-consistent classifications across views. 

Once the base orientation is fixed, we generate a small set of discrete orientation hypotheses that account for possible viewpoint ambiguities. Intuitively, this step handles cases where the rendered views correspond to the same physical orientation but appear in a different order due to camera sampling.

For example, suppose the four rendered views of $\basemesh$ correspond to the sequence
\texttt{[front, left, back, right]}. If the base object is rotated by $90^\circ$ around its vertical axis, the same viewpoints would instead appear as
\texttt{[left, back, right, front]}. We therefore consider all such cyclic shifts of the rendered view sequence to account for these azimuthal offsets. To handle this case where some objects appear upside down in the image, we also include a hypothesis that applies a $180^\circ$ in-plane rotation, which reverses the predicted up direction of the object. In total, there are 8 such possible cyclic shifts that account for the mismatch in orientation, all of which we consider as our hypothesis.

For each candidate orientation of $\objectmesh$, we render $M$ views and classify them using the scorer. The scorer predicts the labels \texttt{visible\_face} and \texttt{projected\_up} for each rendered image together with confidence values for each prediction.

To evaluate a candidate orientation, we compare these predicted labels against those of $\basemesh$ under each of the discrete orientation hypotheses described above. For every view, we check whether the predicted side of the object (e.g., front, left) and the predicted upward direction match those of the base object. A match contributes positively to the score, and this contribution is weighted by the confidence of the corresponding prediction.

The contributions from all views are then averaged to produce a score for the current hypothesis. In addition, we apply a penalty when the sequence of predicted viewpoints across the rendered views is geometrically inconsistent (for example, if the sequence viewpoints do not follow a plausible progression around the object). The final score for a candidate orientation of $\objectmesh$ is then the maximum score obtained across all orientation hypotheses.

Intuitively, this scoring procedure favors orientations where the target object consistently appears aligned with the base object across multiple viewpoints. Orientations that produce mismatched sides, inconsistent upward directions, or implausible viewpoint sequences receive lower scores. Now among $N=20$ orientation of the $\objectmesh$, we simply select the orientation that achieves the highest score as the final initialization.

The final, initialized target mesh is then obtained by applying a similarity transform to the original target vertices, using the orientation selected as giving the highest score:
\[
\mathbf{v}_i^{\,0}
=
s^* \, R^* \bigl(\mathbf{v}_i - \bar{\mathbf{v}}\bigr) + \mathbf{t}^* ,
\]
where \(\mathbf{v}_i\) is a vertex of the original target mesh, \(\bar{\mathbf{v}}=\frac{1}{N}\sum_i \mathbf{v}_i\) is the target-mesh centroid, \(R^*\) is the final rotation, \(s^*\) is the scale, and \(\mathbf{t}^*\) is the translation. In our implementation,
\[
R^* = (R_b^*)^\top R_h^* R_t^*,
\]
where \(R_t^*\) is the best target rotation found during VLM search, \(R_b^*\) is the rotation initially applied to the base mesh to find the best canonical rotation, and \(R_h^*\) is the best hypothesis correction (the cyclic shift we applied, if theres any). The translation is set to
\[
\mathbf{t}^* = \mathbf{c}_{\text{region}} + \boldsymbol{\eta},
\]
where \(\mathbf{c}_{\text{region}}\) is the centroid of the user-defined highlighted region and \(\boldsymbol{\eta}\) is an optional Gaussian jitter term. The scale is computed as
\[
s^* = 2.0 \cdot
\frac{\left\|\mathbf{b}^{\max}_{\text{region}}-\mathbf{b}^{\min}_{\text{region}}\right\|_2}
{\left\|\mathbf{b}^{\max}_{\text{target}}-\mathbf{b}^{\min}_{\text{target}}\right\|_2+\varepsilon}.
\]

\paragraph{Material Regularization.}
In this section, we provide more details of how we formulate the material regularization loss.
Recall that we use a hyperelastic material model to compute a \texttt{Neo-Hookean} loss function that models the internal forces resisting deformation from each jacobians, $J_i$. 
The function takes the deformation gradient $F$ (which we discretely approximate using our jacobians) for each element, along with two material properties: Young's modulus $E$ and Poisson's ratio $\nu$ (predicted from our \textit{Material Verifier} agent). These are first converted into Lamé parameters: the shear modulus $\mu$ and Lamé's first parameter $\lambda$.

$$
\mu = \frac{E}{2(1 + \nu)} \quad \text{and} \quad \lambda = \frac{E\nu}{(1 + \nu)(1 - 2\nu)}
$$
The function then computes the first Piola-Kirchhoff stress tensor $P$, which represents the internal forces for each element:
$$
P(F) = \mu (F - F^{-T}) + \lambda \ln(\det(F)) F^{-T}
$$
where $\det(F)$ is the volumetric determinant of the deformation gradient. To apply this continuous model to our discrete mesh, we use our piece-wise Jacobians, $J_i$ to serve as our replacement for the deformation gradient $F$, so the stress for each face can be approximated as, 
$$
P(J_i) = \mu (J_i - J_i^{-T}) + \lambda \ln(\det(J_i)) J_i^{-T}
$$
and subsequently, 

$$
\mathcal{L}_{e} =  \sum_i  P(J_i).
$$
We defer more details to \cite{sifakis2012fem}.

As a result of this deformation, we can obtain a deformed object mesh that tightly and semantically fits around the user-defined region on the base mesh, while strictly abiding by the physically-imposed requirement of non-intersection. We also note that the $E$ value under 1000 makes the $\objectmesh$ fit too elastically over the $\basemesh$, up to a point where it sacrifices semantic realism. We therefore clamp the minimum value of $E$ to 1000 for the purpose of our experiments, which, with an iterative deformation process, successfully models even the most elastic materials we show in our experiments (e.g. ``extremely soft cloth").

\section{Additional Experiments and Results}
\label{sec:additional_results}
\subsection{User Study}
\label{subsec:user_study}

We conduct an extensive user study for 32 participants, using the same 10 set of examples we used for our quantitative evaluation. For each example, we show 6 randomly ordered results from ours and other baseline methods. We then ask the participants to choose the single best result among the 6 options that ``performs the task of putting on a wearable 3D object in the highest quality possible--in a way that best matches the text description provided above in each section.''
As can be seen from \autoref{tab:user_study}, our method has consistently been ranked by most users to give the most realistic composition result compared to the other 5 baselines. While our method has an overall average preference of 92.81, it also has an extremely strong minimum preference score (the preference score of the example where the method achieved its lowest user preference among all evaluation examples) of 81.25, meaning that even for the ``worst'' example, 81.25 percent of users preferred our results over all the baselines.

\begin{table}[t]
\vspace{4mm}
\centering
\small
\setlength{\tabcolsep}{6pt}
\renewcommand{\arraystretch}{1.2}
\setlength{\aboverulesep}{0pt}
\setlength{\belowrulesep}{0pt}

\resizebox{\columnwidth}{!}{%
\begin{tabular}{lcccccc}
\toprule
\multicolumn{7}{l}{\textbf{User Study Results}}\\
\rowcolor{headercolor}
Statistic & \shortstack{Instant3Dit} & \shortstack{Best ICP} & \shortstack{Best FGR} & \shortstack{RANSAC+ICP} & \shortstack{RANSAC+FGR} & \shortstack{Ours} \\
\midrule
\rowcolor{zebra}
Average Preference (\%) $\uparrow$ & 2.81 & 3.13 & 0.63 & 0.00 & 0.63 & \textbf{92.81} \\
Minimum (\%) & 0.00 & 0.00 & 0.00 & 0.00 & 0.00 & 81.25 \\
\rowcolor{zebra}
Maximum (\%) & 15.63 & 12.50 & 3.13 & 0.00 & 3.13 & \textbf{100.00} \\
\bottomrule
\end{tabular}%
}
\vspace{4mm}
\caption{User study results comparing six methods for the 3D wearable compositing task. Each entry reports the percentage of participant selections received by each method across all evaluation prompts (32 participants per prompt). We report the average preference percentage, as well as the minimum and maximum percentage achieved by each method on any individual example.}
\vspace{-2mm}
\label{tab:user_study}
\end{table}
\vspace{-3mm}
\subsection{Experiment Details}
\label{subsec:experiment_details}
\paragraph{Quantitative Comparison} In this section, we will detail how we conduct our quantitative experiments. Among the 5 baselines we compare against in both our quantitative evaluation and user study, we first discuss how we use Instan3dit. Instant3dit takes as input a source mesh and a 3D mask that denotes the local region to make edits. As their 3D mask is not identical to our 2D surface highlights, we manually generate 3D masks that are as close as possible to our 2D highlight. We also note that these 3D masks are more expressive user inputs than our surface highlights, and thus shoudn't be considered a factor that confers our method a quantitative/qualitative edge. One major caveat of comparing against Instant3dit is that the backbone Large Reconstruction Model (LRM) the authors used for their paper is not publicly available (the Adobe LRM). We therefore resorted to the alternative InstantMesh \cite{xu2024instantmeshefficient3dmesh} backbone suggested in their GitHub repositories to run the experiments, which apparently degrades the quality of the results. 

For quantitative evaluation, we ran Instant3dit from the same text prompt processed by \ourmethod{}'s VLM verifiers and the diffusion model used for deformation guidance. We rendered meshes generated from each method in a uniform, grey textured color from 4 manually sampled viewpoints, and averaged the scores across these viewpoints.

\paragraph{Comparison with Point Registration Methods} We also made qualitative/quantitative comparisons against the Point Cloud Registration Methods, including Iterative Closest Points (ICP), Fast Global Registration (FGR), and Random Sample Consensus (Ransac). We note that all of these methods in a standalone setting is not robust enough (or has too much variance) to meaningfully compare against our method. We therefore take standard practices known to improve these methods' performances, and compare the improved results against our method. For ICP, whose results are known to be greatly influenced by initialization, we sample results from $n=50$ different initializations, and choose the one that gives best RMSE score (we call these \textbf{Best ICP} and \textbf{Best FGR} since we choose the result from best initialiization). Another approach is to use Ransac (which is relatively better at finding coarse global registrations) as an initialization, and using ICP as a second-step, finetuning method (which we call \textbf{Ransac+ICP} and \textbf{Ransac+FGR} in our main paper). We compare our results to both methods in the paper. We also found that for this application, an abnormally high threshold value works best in the most stable, robust manner, so we use $th=5000$ as our ICP threshold. 

For FGR, we take the same approach and run comparisons on both multi-step sampling and Ransac initialization method. We use a voxel size of 0.05 and a distance multiplier of 1.5, which is a default value proposed in its \texttt{Open3d} implementation.

\paragraph{Examples of user study and quantitative comparison}
In Figure (\ref{fig:quant_examples}), we provide examples of the renderings we used for conducting user study and quantitative evaluations. 
\section{Computational Details} 
\label{sec:computational_details}
We ran all of our experiments on single a40 nvidia gpus. 
While the runtime depends heavily on the mesh resolution (and the size of user input mask), our method approximately takes around 15-30 minutes to run. The VLM verification stage alone takes a minimum of 5 minutes to run, causing a major bottleneck to the model. While these performance estimates are that of the results we presented in this paper, the total runtime could be greatly reduced without sacrificing too much quality by running each step for significantly less iterations. We will release the exact parameters and code upon acceptance. \paragraph{GPU-Optimized BVH memory and runtime test}
In Figure (\ref{fig:bvh_quant}), we provide an extensive experiment of varying the $\texttt{leaf\_size}$ parameter of our gpu-optimized BVH structure in calculating the vertex-to-triangle distance measure $\hat{d}(v_i,\mathcal{M}_b)$, as well as doing a single fully-batched computation. Specifically, we calculate $\hat{d}(v_i,\mathcal{M}_b)$ using bvh with varying $\texttt{leaf\_size}$, applied onto progressively subdivided mesh resolutions. For $\objectmesh$, we choose a reasonable resolution that is the approximately the average of object meshes we used in our experiments, with 8646 vertices and 17288 faces.
We show that  $\texttt{leaf\_size=1000}$ provides a reasonable trade-off between runtime and GPU memory, while running the fully-batched computation without using our BVH structure crashes on high-resolution meshes. We note that the fully-batched computation crashes sooner for lower resolution triangles when employed in our actual optimization pipeline, where factors like autograd trees and backpropagation further constrain gpu memory.
\newpage

\begin{center}
    \includegraphics[width=\textwidth]{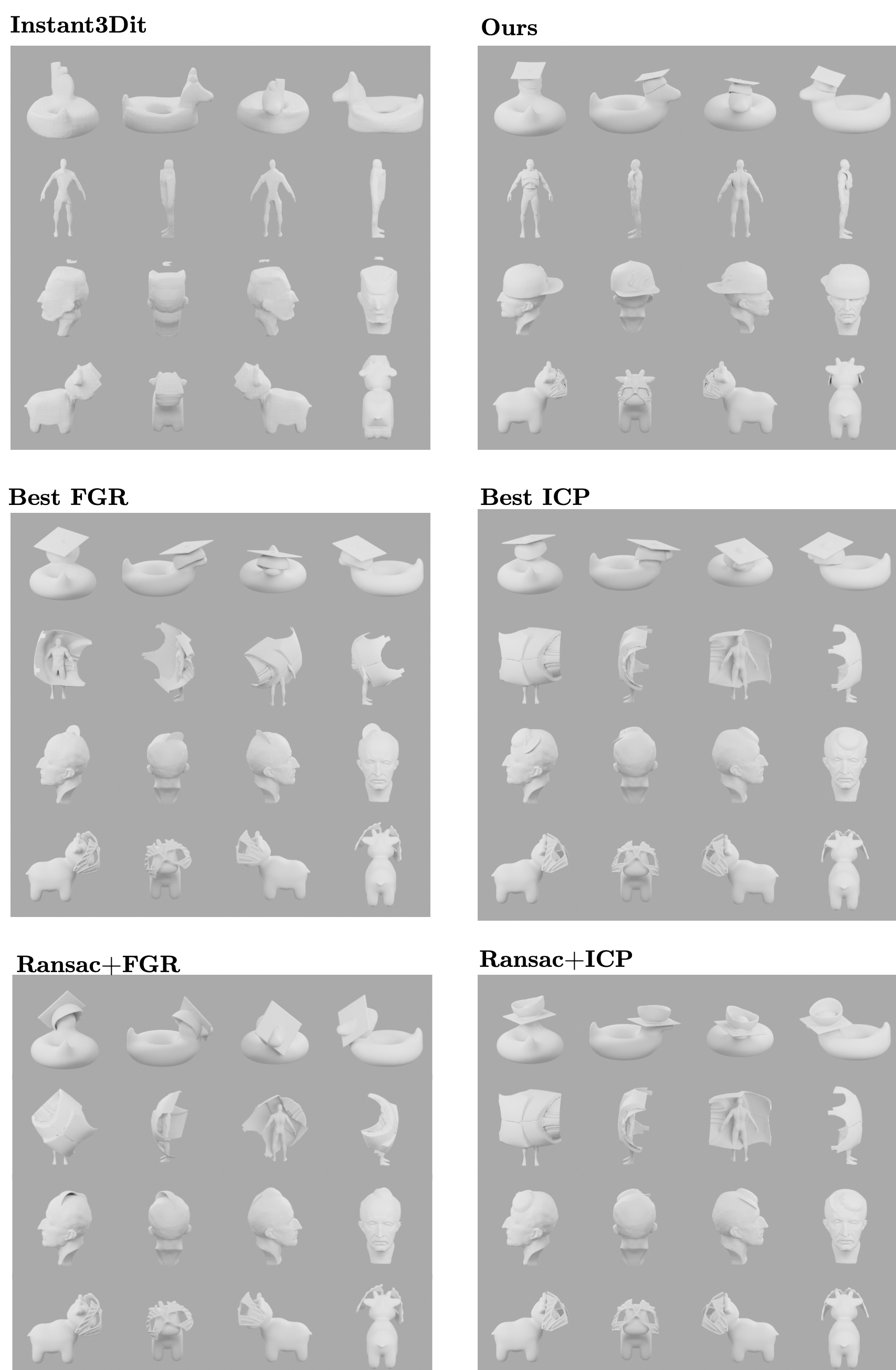}
    \vspace{-0.7cm}
    \captionof{figure}{We show examples of renderings that we used for our user study. We also used the same set of examples for evaluating CLIP, CLIP-IQA, and VQA scores, but with a different renderer (nvdiffrast) for technical purposes. We render results from all methods from four viewpoints with greyscale textures.}
    \label{fig:quant_examples}
\end{center}

\newcommand{\placeholder}[1]{\texttt{\textcolor{blue}{<#1>}}}

\section{VLM Prompt Templates}

\begin{center}
\small
\captionof{table}{\textbf{Prompt for Agent 1: Semantic Describer.} This agent generates an immutable object-centric canonical reference frame, concise view definitions for both objects, and a material description for the target object.}
\label{tab:prompt_describer}
\begin{tabularx}{\linewidth}{X}
\toprule
\textbf{System Instruction} \\
\midrule
You are an expert in 3D object semantics, pose, and materials. You MUST define an immutable object-centric canonical frame that will be reused later for view classification. This frame must be stable across renders and must not depend on a specific camera. \\

\textbf{OBJECTS:} \\
- \textbf{TARGET:} \placeholder{target\_label} \\
- \textbf{BASE:} \placeholder{base\_label} \\
- \textbf{VISUALS:} [Optional rendered images of target and base meshes from multiple turntable views] \\

\textbf{PART 1: OBJECT-CENTRIC CANONICAL FRAMES} \\
For each object, define an object-centric canonical frame, including: \\
- an origin landmark, \\
- semantic $+X$ (right), \\
- semantic $+Y$ (up), \\
- semantic $+Z$ (front / outward-facing direction), \\
- at least three cues for distinguishing left vs.\ right, \\
- at least three cues for distinguishing front vs.\ back, \\
- at least two cues for distinguishing top vs.\ bottom. \\

IMPORTANT: The target frame should be defined relative to how the target is functionally worn on the base, but the axes themselves must remain object-centric and reusable across views. \\

\textbf{PART 2: REQUIRED CLASSIFIER LABELS} \\
A later classifier will see images of only one object at a time. For each image it must output: \\
- \texttt{visible\_face} $\in$ \{\texttt{front}, \texttt{back}, \texttt{left}, \texttt{right}, \texttt{top}, \texttt{bottom}, \texttt{other}\}, indicating the dominant canonical side facing the camera; \\
- \texttt{projected\_up} $\in$ \{\texttt{up}, \texttt{down}, \texttt{left}, \texttt{right}, \texttt{ambiguous}\}, indicating the image-plane direction of the object's canonical $+Y$ axis. \\

Provide concise definitions for these labels for both the target and the base. \\

\textbf{PART 3: MATERIAL DESCRIPTION} \\
Describe the physical material properties of the \textbf{TARGET object only}. This natural-language description will later be converted into approximate simulation parameters. \\

\textbf{OUTPUT FORMAT:} \\
You MUST provide your response only as a single valid JSON object with four keys: \\
\texttt{canonical\_reference}, \texttt{target\_view\_definitions}, \texttt{base\_view\_definitions}, and \texttt{material\_description}. \\
\bottomrule
\end{tabularx}
\end{center}

\vspace{1em}
\newpage
\begin{center}
\small
\captionof{table}{\textbf{Prompt for Agent 2: View Classifier (Scorer).} This agent classifies views of a \emph{single} object at a time using the immutable object-centric canonical frame generated by Agent 1.}
\label{tab:prompt_scorer}
\begin{tabularx}{\linewidth}{X}
\toprule
\textbf{System Instruction} \\
\midrule
You are a strict viewpoint classifier. You will be shown images of only one object at a time, either the TARGET or the BASE. You must use the immutable object-centric canonical frame below and must not redefine axes for each image. \\

\textbf{INPUT CONTEXT:} \\
- \textbf{OBJECT ROLE:} \placeholder{TARGET or BASE} \\
- \textbf{OBJECT LABEL:} \placeholder{object\_label} \\
- \textbf{OBJECT-CENTRIC CANONICAL FRAME:} \placeholder{canonical\_reference (from Agent 1)} \\
- \textbf{CONCISE DEFINITIONS FOR THIS OBJECT:} \placeholder{target/base view definitions (from Agent 1)} \\
- \textbf{VISUALS:} [Rendered images of the object from multiple viewpoints] \\

\textbf{TASK:} \\
For each image, output: \\
- \texttt{visible\_face} $\in$ \{\texttt{front}, \texttt{back}, \texttt{left}, \texttt{right}, \texttt{top}, \texttt{bottom}, \texttt{other}\}, indicating the dominant canonical side facing the camera; \\
- \texttt{projected\_up} $\in$ \{\texttt{up}, \texttt{down}, \texttt{left}, \texttt{right}, \texttt{ambiguous}\}, indicating the image-plane direction of the object's canonical $+Y$ axis; \\
- \texttt{face\_confidence} $\in [0,1]$; \\
- \texttt{up\_confidence} $\in [0,1]$; \\
- a short textual reason citing one or two concrete visual cues. \\

\textbf{IMPORTANT RULES:} \\
- Do not mention any other object. \\
- Do not try to force agreement with any external reference. \\
- Use only the supplied canonical frame and visual evidence. \\
- Output only JSON. \\

\textbf{OUTPUT FORMAT:} \\
Respond only with a single JSON object containing a list named \texttt{view\_analysis}. Each entry should contain: \\
\texttt{view\_pair}, \texttt{visible\_face}, \texttt{projected\_up}, \texttt{face\_confidence}, \texttt{up\_confidence}, and \texttt{reason}. \\
\bottomrule
\end{tabularx}
\end{center}

\newpage
\begin{center}
\small
\captionof{table}{\textbf{Prompt for Agent 3: Material Verifier.} This agent converts the natural-language material description produced by Agent 1 into approximate elastic parameters for simulation.}
\label{tab:prompt_material}
\begin{tabularx}{\linewidth}{X}
\toprule
\textbf{System Instruction} \\
\midrule
You are a materials expert. Based on the following description of an object, infer approximate elastic parameters for simulation. \\

\textbf{INPUT:} \\
Infer simulation parameters from this description of a '\placeholder{target\_label}': '\placeholder{material\_description (from Agent 1)}'. \\

\textbf{OUTPUT FORMAT:} \\
Output strictly in JSON with keys: \{\texttt{E}, \texttt{nu}, \texttt{mu}\}, corresponding to Young's modulus, Poisson's ratio, and shear modulus. \\

\textbf{GUIDELINES:} \\
- Units for \texttt{E} and \texttt{mu} are arbitrary but should be proportional across materials. \\
- \texttt{nu} should lie in $[0, 0.5]$. \\
- For technical stability, \texttt{E} must be at least $1000$; if the inferred value is smaller, clamp it to $1000$. \\
\bottomrule
\end{tabularx}
\end{center}
\newpage

\end{document}